\documentclass[%
    reprint,
    superscriptaddress,
    showkeys,
    nofootinbib, nobibnotes,
    amsmath, amssymb,
    aps, prd,
    floatfix,
    hyphens, spaces, obeyspaces
]{revtex4-2}

\usepackage{dcolumn}
\usepackage{bm}
\usepackage[english]{babel}
\usepackage[utf8]{inputenc}
\usepackage[T1]{fontenc}
\usepackage{soul}
\usepackage{color}
\usepackage[table,dvipsnames]{xcolor}
\definecolor{myblue}{RGB}{13, 71, 161}
\definecolor{mygreen}{RGB}{115, 140, 84}
\definecolor{myred}{RGB}{216, 28, 56}
\definecolor{myblack}{RGB}{0, 0, 0}
\definecolor{mybrown}{RGB}{102, 80, 10}
\usepackage[
    colorlinks=true,
    breaklinks=true,
    hyperfootnotes=true,
    urlcolor=myblue,
    linkcolor=myblue,
    citecolor=OliveGreen,
]{hyperref}
\usepackage[nameinlink]{cleveref}
\usepackage{nameref}
\usepackage{booktabs}       %
\usepackage{amsfonts}       %
\usepackage{nicefrac}       %
\usepackage{microtype}      %
\usepackage{lipsum}
\usepackage{graphicx}

\graphicspath{{Fig/}}

\Crefname{section}{Sec.}{Sec.}
\Crefname{figure}{Fig.}{Figs.}
\Crefname{table}{Tab.}{Tabs.}

\usepackage{tikz}
\usepackage[edges]{forest}

\definecolor{paired-light-blue}{RGB}{198, 219, 239}
\definecolor{paired-dark-blue}{RGB}{49, 130, 188}
\definecolor{paired-light-orange}{RGB}{251, 208, 162}
\definecolor{paired-dark-orange}{RGB}{230, 85, 12}
\definecolor{paired-light-green}{RGB}{199, 233, 193}
\definecolor{paired-dark-green}{RGB}{49, 163, 83}
\definecolor{paired-light-purple}{RGB}{218, 218, 235}
\definecolor{paired-dark-purple}{RGB}{117, 107, 176}
\definecolor{paired-light-gray}{RGB}{217, 217, 217}
\definecolor{paired-dark-gray}{RGB}{99, 99, 99}
\definecolor{paired-light-pink}{RGB}{222, 158, 214}
\definecolor{paired-dark-pink}{RGB}{123, 65, 115}
\definecolor{paired-light-red}{RGB}{231, 150, 156}
\definecolor{paired-dark-red}{RGB}{131, 60, 56}
\definecolor{paired-light-yellow}{RGB}{231, 204, 149}
\definecolor{paired-dark-yellow}{RGB}{141, 109, 49}
\tikzset{%
    parent/.style =          {align=center,text width=1.5cm,rounded corners=3pt, line width=0.3mm, fill=gray!10,draw=gray!80},
    child/.style =           {align=center,text width=2.3cm,rounded corners=3pt, fill=blue!10,draw=blue!80,line width=0.3mm},
    grandchild/.style =      {align=center,text width=2cm,rounded corners=3pt},
    greatgrandchild/.style = {align=center,text width=1.5cm,rounded corners=3pt},
    greatgrandchild2/.style = {align=center,text width=1.5cm,rounded corners=3pt},
    referenceblock/.style =  {align=center,text width=1.5cm,rounded corners=2pt},
    top_class/.style =           {align=center,text width=2cm,rounded corners=3pt, fill=paired-light-gray!50,draw=paired-dark-gray!65,line width=0.3mm},
    generation/.style =           {align=center,text width=2cm,rounded corners=3pt, fill= paired-light-green!50,draw=paired-dark-green!75,line width=0.3mm},
    generation_wide/.style =           {align=center,text width=2.5cm,rounded corners=3pt, fill= paired-light-green!50,draw=paired-dark-green!75,line width=0.3mm},
    generation_more/.style =           {align=center,text width=4cm,rounded corners=3pt, fill= paired-light-green!50,draw=paired-dark-green!75,line width=0.3mm},
    generation_work/.style =           {align=center,text width=4.0cm,rounded corners=3pt, fill= paired-light-green!50,draw= cyan!0,line width=0.3mm},
    encoder/.style =           {align=center,text width=2cm,rounded corners=3pt, fill=paired-light-orange!50,draw=paired-dark-orange!65,line width=0.3mm},
    encoder_more/.style =           {align=center,text width=4cm,rounded corners=3pt, fill=paired-light-orange!50,draw=paired-dark-orange!65,line width=0.3mm},
    encoder_work/.style =           {align=center,text width=4.0cm,rounded corners=3pt, fill=paired-light-orange!50,draw=red!0,line width=0.3mm},
    gpa/.style =           {align=center,text width=2cm,rounded corners=3pt, fill=paired-light-blue!50,draw=paired-dark-blue!65,line width=0.3mm},
    gpa_wide/.style =           {align=center,text width=4cm,rounded corners=3pt, fill=paired-light-blue!50,draw=paired-dark-blue!65,line width=0.3mm},
    gpa_work/.style =           {align=center, text width=4.0cm,rounded corners=3pt, fill=paired-light-blue!50,draw=blue!0,line width=0.3mm},
    data/.style =           {align=center,text width=2cm,rounded corners=3pt, fill=paired-light-blue!50,draw=paired-dark-blue!65,line width=0.3mm},
    data_wide/.style =           {align=center,text width=3cm,rounded corners=3pt, fill=paired-light-blue!50,draw=paired-dark-blue!65,line width=0.3mm},
    data_work/.style =           {align=center, text width=4.5cm,rounded corners=3pt, fill=paired-light-blue!50,draw=blue!0,line width=0.3mm},
    model/.style =           {align=center,text width=2cm,rounded corners=3pt, fill=paired-light-orange!50,draw=paired-dark-orange!65,line width=0.3mm},
    model_more/.style =           {align=center,text width=4cm,rounded corners=3pt, fill=paired-light-orange!50,draw=paired-dark-orange!65,line width=0.3mm},
    model_work/.style =           {align=center,text width=4.5cm,rounded corners=3pt, fill=paired-light-orange!50,draw=red!0,line width=0.3mm},
    pretraining/.style =           {align=center,text width=2cm,rounded corners=3pt, fill= paired-light-green!50,draw=paired-dark-green!75,line width=0.3mm},
    pretraining_wide/.style =           {align=center,text width=2.5cm,rounded corners=3pt, fill= paired-light-green!50,draw=paired-dark-green!75,line width=0.3mm},
    pretraining_more/.style =           {align=center,text width=4cm,rounded corners=3pt, fill= paired-light-green!50,draw=paired-dark-green!75,line width=0.3mm},
    pretraining_work/.style =           {align=center,text width=4.5cm,rounded corners=3pt, fill= paired-light-green!50,draw= cyan!0,line width=0.3mm},
    finetuning/.style =           {align=center,text width=2cm,rounded corners=3pt, fill= paired-light-purple!50,draw=paired-dark-purple!75,line width=0.3mm},
    finetuning_work/.style =           {align=center,text width=4.5cm,rounded corners=3pt, fill= paired-light-purple!50,draw= orange!0,line width=0.3mm},
    inference/.style =           {align=center,text width=2cm,rounded corners=3pt, fill= paired-light-red!35,draw=paired-light-red!90,line width=0.3mm},
    inference_more/.style =           {align=center,text width=4cm,rounded corners=3pt, fill= paired-light-red!35,draw=paired-light-red!90,line width=0.3mm},
    inference_work/.style =           {align=center,text width=4.5cm,rounded corners=3pt, fill= paired-light-red!35,draw= magenta!0,line width=0.3mm},
}

\usepackage{acro}
\acsetup{make-links=true}

\DeclareAcroEnding{possessive}{'s}{'s}
\NewAcroCommand\acg{m}{\acropossessive\UseAcroTemplate{first}{#1}}
\NewAcroCommand\acsg{m}{\acropossessive\UseAcroTemplate{short}{#1}}
\NewAcroCommand\aclg{m}{\acropossessive\UseAcroTemplate{long}{#1}}
\NewAcroCommand\acfg{m}{%
    \acrofull
    \acropossessive
    \UseAcroTemplate{first}{#1}%
}
\NewAcroCommand\iacsg{m}{%
    \acroindefinite
    \acropossessive
    \UseAcroTemplate{short}{#1}%
}

\DeclareAcronym{AGN}{
    short = AGN,
    long  = Active Galactic Nuclei
}

\DeclareAcronym{AI}{
    short = AI,
    long  = artificial intelligence
}
\DeclareAcronym{AIGC}{
    short = AIGC,
    long  = artificial intelligence-generated content
}

\DeclareAcronym{AK}{
    short = AK,
    long  = ``analytic kludge"
}
\DeclareAcronym{AAK}{
    short = AAK,
    long  = argumented analytic kludge
}
\DeclareAcronym{AUC}{
    short = AUC,
    long  = area under the curve
}
\DeclareAcronym{BH}{
    short = BH,
    long  = black hole
}
\DeclareAcronym{BHB}{
    short = BHB,
    long  = black hole binary,
    long-plural-form = black hole binaries
}

\DeclareAcronym{BBH}{
    short = BBH,
    long  = binary black hole
}

\DeclareAcronym{BNS}{
    short = BNS,
    long  = neutron star
}

\DeclareAcronym{CCSN}{
    short = CCSN,
    long  = core-collapse supernovae
}

\DeclareAcronym{CNN}{
    short = CNN,
    long  = convolutional neural network
}

\DeclareAcronym{DECIGO}{
    short = DECIGO,
    long  = DECi-hertz Interferometer Gravitational wave Observatory
}

\DeclareAcronym{DL}{
    short = DL,
    long  = deep learning
}

\DeclareAcronym{DNN}{
    short = DNN,
    long  = deep neural network
}
\DeclareAcronym{EoS}{
    short = EoS,
    long  = equation of state
}
\DeclareAcronym{ESA}{
    short = ESA,
    long  = European Space Agency
}
\DeclareAcronym{EMRI}{
    short = EMRI,
    long  = extreme-mass-ratio inspiral
}

\DeclareAcronym{ET}{
    short = ET,
    long  = Einstein Telescope
}

\DeclareAcronym{FAR}{
    short = FAR,
    long  = false alarm rate
}

\DeclareAcronym{FPR}{
    short = FPR,
    long  = false positive rate
}

\DeclareAcronym{GAN}{
    short = GAN,
    long  = generative adversarial network
}

\DeclareAcronym{GB}{
    short = GB,
    long  = galactic binary,
    long-plural-form = galactic binaries
}

\DeclareAcronym{GP}{
    short = GP,
    long  = Gaussian Processes,
}

\DeclareAcronym{GR}{
    short = GR,
    long  = general relativity
}
\DeclareAcronym{GW}{
    short = GW,
    long  = gravitational wave
}

\DeclareAcronym{GWDA}{
    short = GWDA,
    long  = gravitational wave data analysis
}
\DeclareAcronym{LDC}{
    short = LDC,
    long  = LISA Data Challenge
}

\DeclareAcronym{LIGO}{
    short = LIGO,
    long  = \href{http://www.ligo.caltech.edu/}{Laser Interferemeter Gravitational Wave Observatory}
}
\DeclareAcronym{LISA}{
    short = LISA,
    long  = \href{https://www.lisamission.org/}{Laser Interferometer Space Antenna}
}

\DeclareAcronym{LLM}{
    short = LLM,
    long  = large language model
}

\DeclareAcronym{MBH}{
    short = MBH,
    long  = massive black hole
}

\DeclareAcronym{MBHB}{
    short = MBHB,
    long  = massive black hole binary,
    long-plural-form = massive black hole binaries
}

\DeclareAcronym{MCMC}{
    short = MCMC,
    long  = Markov-chain Monte Carlo
}

\DeclareAcronym{MLDC}{
    short = MLDC,
    long  = \href{http://astrogravs.nasa.gov/docs/mldc/}{Mock LISA Data Challenge}
}

\DeclareAcronym{NK}{
    short = NK,
    long  = numerical kludge
}

\DeclareAcronym{NR}{
    short = NR,
    long  = numerical relativity
}

\DeclareAcronym{OMS}{
    short = OMS,
    long  = optical metrology system
}

\DeclareAcronym{PE}{
    short = PE,
    long  = parameter estimation
}

\DeclareAcronym{PSD}{
    short = PSD,
    long  = power spectral density
}
\DeclareAcronym{ReLU}{
    short = ReLU,
    long  = rectified linear
    unit
}

\DeclareAcronym{ResNet}{
    short = ResNet,
    long  = residual network
}

\DeclareAcronym{RLHF}{
    short = RLHF,
    long  = reinforcement learning from human feedback
}

\DeclareAcronym{RNN}{
    short = RNN,
    long  = recurrent neural network
}

\DeclareAcronym{ROC}{
    short = ROC,
    long  = receiver operating characteristic
}

\DeclareAcronym{SGWB}{
    short = SGWB,
    long  = stochastic gravitational wave background
}

\DeclareAcronym{SMBH}{
    short = SMBH,
    long  = super-massive black hole
}

\DeclareAcronym{SNR}{
    short = SNR,
    long  = signal-to-noise ratio
}

\DeclareAcronym{SOBH}{
    short = SOBH,
    long  = stellar origin black hole binary
}

\DeclareAcronym{SSB}{
    short = SSB,
    long  = solar system barycenter
}

\DeclareAcronym{TDI}{
    short = TDI,
    long  = time delay interferometry
}

\DeclareAcronym{TPR}{
    short = TPR,
    long  = true positive rate
}

\DeclareAcronym{t-SNE}{
    short = t-SNE,
    long  = t-distributed stochastic neighbor embedding
}

\DeclareAcronym{VAE}{
    short = VAE,
    long  = variational autoencoder,
}

\DeclareAcronym{VGB}{
    short = VGB,
    long  = verification galactic binary,
    long-plural-form = verification galactic binaries
}

\begin{document}

\preprint{APS/123-QED}

\title{Dawning of a New Era in Gravitational Wave Data Analysis: \\ Unveiling Cosmic Mysteries via Artificial Intelligence --- A Systematic Review}

\author{Tianyu Zhao}
\affiliation{Center for Gravitational Wave Experiment, National Microgravity Laboratory, Institute of Mechanics, Chinese Academy of Sciences, Beijing 100190, China}
\affiliation{School of Physics and Astronomy, Beijing Normal University, Beijing 100875, China}
\affiliation{Peng Cheng Laboratory, Shenzhen, 518055, China}
\affiliation{Institute for Frontiers in Astronomy and Astrophysics, Beijing Normal University, Beijing 102206, China}

\author{Ruijun Shi}%
\affiliation{School of Physics and Astronomy, Beijing Normal University, Beijing 100875, China}
\affiliation{Institute for Frontiers in Astronomy and Astrophysics, Beijing Normal University, Beijing 102206, China}

\author{Yue Zhou}
\affiliation{Peng Cheng Laboratory, Shenzhen, 518055, China}

\author{Zhoujian Cao}%
\thanks{Corresponding author: \href{mailto:zjcao@bnu.edu.cn}{zjcao@bnu.edu.cn}}
\affiliation{School of Physics and Astronomy, Beijing Normal University, Beijing 100875, China}
\affiliation{Institute for Frontiers in Astronomy and Astrophysics, Beijing Normal University, Beijing 102206, China}
\affiliation{School of Fundamental Physics and Mathematical Sciences, Hangzhou Institute for Advanced Study, UCAS, Hangzhou 310024, China}

\author{Zhixiang Ren}
\thanks{Corresponding author: \href{mailto:renzhx@pcl.ac.cn}{renzhx@pcl.ac.cn}}
\affiliation{Peng Cheng Laboratory, Shenzhen, 518055, China}

\date{\today}%

\begin{abstract}
	\begin{minipage}[t]{0.46\textwidth}
		Gravitational wave data analysis (GWDA) faces significant challenges due to high-dimensional parameter spaces and non-Gaussian, non-stationary artifacts in the interferometer background, which traditional methods have made significant progress in addressing but continue to face limitations. Artificial intelligence (AI), particularly deep learning (DL) algorithms, offers potential advantages, including computational efficiency, scalability, and adaptability, which may complement traditional approaches in tackling these challenges more effectively. In this review, we explore AI-driven approaches to GWDA, covering every stage of the pipeline and presenting first explorations in waveform modeling and parameter estimation. This work represents the most comprehensive review to date, integrating the latest AI advancements with practical GWDA applications. Our meta-analysis reveals insights and trends, highlighting the transformative potential of AI in revolutionizing gravitational wave research and paving the way for future discoveries.
	\end{minipage}
	\hspace{1ex}
	\vtop{%
		\vskip-1ex
		\hbox{%
			\includegraphics[width=0.31\textwidth]{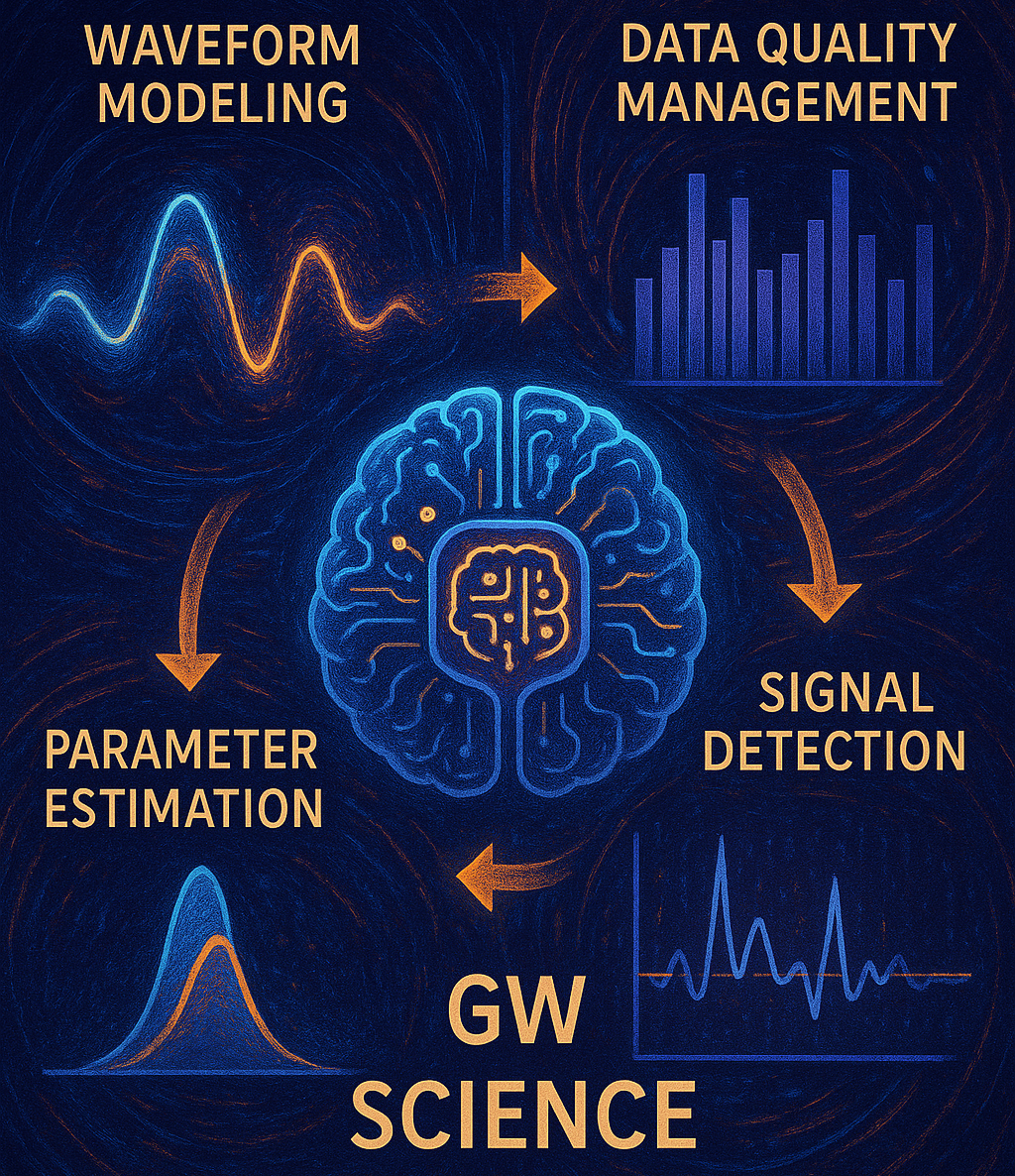}
		}%
	}%
	% \begin{minipage}[t]{0.35\textwidth}
	% 	\includegraphics[width=0.35\textwidth]{graph_abs.png}
	% \end{minipage}
\end{abstract}


\maketitle

\tableofcontents

\section{Introduction}
\label{sec:intro}

In today's rapidly evolving world, \ac{AI} stands as a beacon of
transformation, seamlessly integrating into every facet of our lives. From the
convenience of face recognition \cite{Schroff_2015_CVPR} unlocking our devices
to the magic of speech synthesis \cite{seamlessm4t2023} that brings virtual
assistants to life, \acg{AI} influence is omnipresent. Diving deeper into the
world of language, \acp{LLM} like ChatGPT (GPT-3.5 and GPT-4)
\cite{openai_gpt-4_2023} and Llama 2 \cite{touvron_llama_2023} have
revolutionized our interaction with digital content. These advanced language
models are now being harnessed for creative endeavors, such as co-writing
novels, generating art-inspired poetry \cite{stinger_ai_2023}, and even
composing music \cite{copet_simple_2023}. They're also playing pivotal roles in
bridging communication gaps, offering real-time translations for diplomats and
global travelers, and assisting in preserving endangered languages by
generating rich linguistic content \cite{openai_gpt-4_2023}. Beyond these
linguistic marvels, \acg{AI} imprint extends to personalized healthcare, where
it crafts treatments tailored to individual genetic profiles
\cite{eysenbach_role_2023}, and to our cities, optimizing the ebb and flow of
traffic \cite{djenouri_hybrid_2023}. Venturing into the domain of scientific
inquiry, \ac{AI} emerges as a powerful ally, reshaping our methodologies and
accelerating the pace of discovery \cite{wang_scientific_2023}. Innovations
like self-supervised learning are reimagining how we interpret complex datasets
\cite{olshausen_emergence_1996}, while generative \ac{AI} techniques are
forging new pathways in diverse fields, from drug design
\cite{theodoris_transfer_2023} to the creation of advanced materials
\cite{suwardi_machine_2022}. Delving into specific achievements, tools like
AlphaFold2 \cite{jumper_highly_2021} have unraveled the intricate puzzle of
protein folding, offering unprecedented insights into the very fabric of
biological life. Similarly, in the vast cosmos, \ac{AI} assists astrophysicists
in deciphering the myriad signals from the universe, unveiling celestial
phenomena that were once shrouded in mystery
\cite{bambi_advances_2021}.
As we journey further into this decade, \ac{AI} promises to be more than just a
technological marvel—it is poised to be the compass guiding our quest for
knowledge, reshaping our world, and expanding the horizons of what's possible
\cite{lucci_artificial_2022}.

In 2015, the detection of \acp{GW} provided a monumental breakthrough in
astrophysics \cite{abbott_observation_2016}, validating Einstein's century-old
theoretical prediction \cite{abbott_tests_2016} and introducing a new window to
probe the universe's mysteries
\cite{abbott_gw150914_2016,gonzalez_gravitational_2013}. \ac{GWDA} is a complex endeavor that consist of many stages.
Given the sensitivity of detectors like the \ac{LIGO}
\cite{the_ligo_scientific_collaboration_2015}, Virgo \cite{acernese_2015} and KAGRA \cite{kagra_collaboration_kagra_2019},
it is imperative to differentiate genuine \ac{GW} signals from terrestrial
interference \cite{martynov_sensitivity_2016}. This involves rigorous data
quality labeling \cite{davis_2021}, glitch classification to categorize
transient noise events \cite{mogushi_reduction_2021}, and noise suppression
techniques to enhance the clarity of potential signals
\cite{matichard_seismic_2015}. In addition to these ground-based efforts, upcoming space-based observatories like \ac{LISA}
\cite{amaro-seoane_laser_2017}, Taiji \cite{hu_taiji_2017,ren_taiji_2023}, and
TianQin \cite{luo_tianqin_2016} will open another detection window by targeting sources such as \ac{MBHB},  \ac{EMRI}, \ac{GB}, and \ac{SGWB}. With cleaner data in hand, the focus then shifts to signal detection. For example, template‐based methods rely on accurate waveform modeling to predict gravitational‐wave signals from sources such as black hole mergers \cite{Zhao_2020} and employ matched filtering to detect these signals \cite{finn_detection_1992}. However, signal detection also encompasses template‐free approaches—such as excess power searches and coherent analyses for \acp{SGWB} and unmodeled signals—that identify statistically significant features in the data without relying on precomputed templates \cite{karnesis_assessing_2020}. Following a successful detection, the pipeline
culminates in parameter estimation, where the goal is to decipher the
astrophysical properties of the \ac{GW} sources
\cite{poisson_gravitational_1995}. Each stage of the \ac{GWDA} process, from
data pre-processing to scientific discoveries, presents its unique set of
challenges. Real-time analysis is essential to enable timely communication of detections to the external astronomical community, ensuring that follow-up observations and coordinated multi-messenger campaigns can be initiated promptly. Real-time data analysis necessitates maintaining high data quality because it allows for robust and reliable prompt detections \cite{caballero-garcia_astrophysical_2024,essick_idq_2020}. When it comes to signal
detection, the non-stationary and non-Gaussian nature of the noise poses
significant hurdle, as traditional matched filters are optimized for Gaussian
noise \cite{jaranowski_gravitational-wave_2012}. To adopt matched filtering in non-Gaussian cases, advanced techniques like adaptive \ac{PSD} estimation need to be employed \cite{zackay_detecting_2021}. The generation of waveform templates, crucial for signal searching and parameter estimation, is computationally expensive and time-consuming when relying on classical methods \cite{bohe_improved_2017,varma_surrogate_2019}, resulting in full parameter estimation taking several hours \cite{ashton_bilby_2019,dax_real-time_2021}.

\begin{figure*}[t!]
	\centering
	\includegraphics[width=\textwidth]{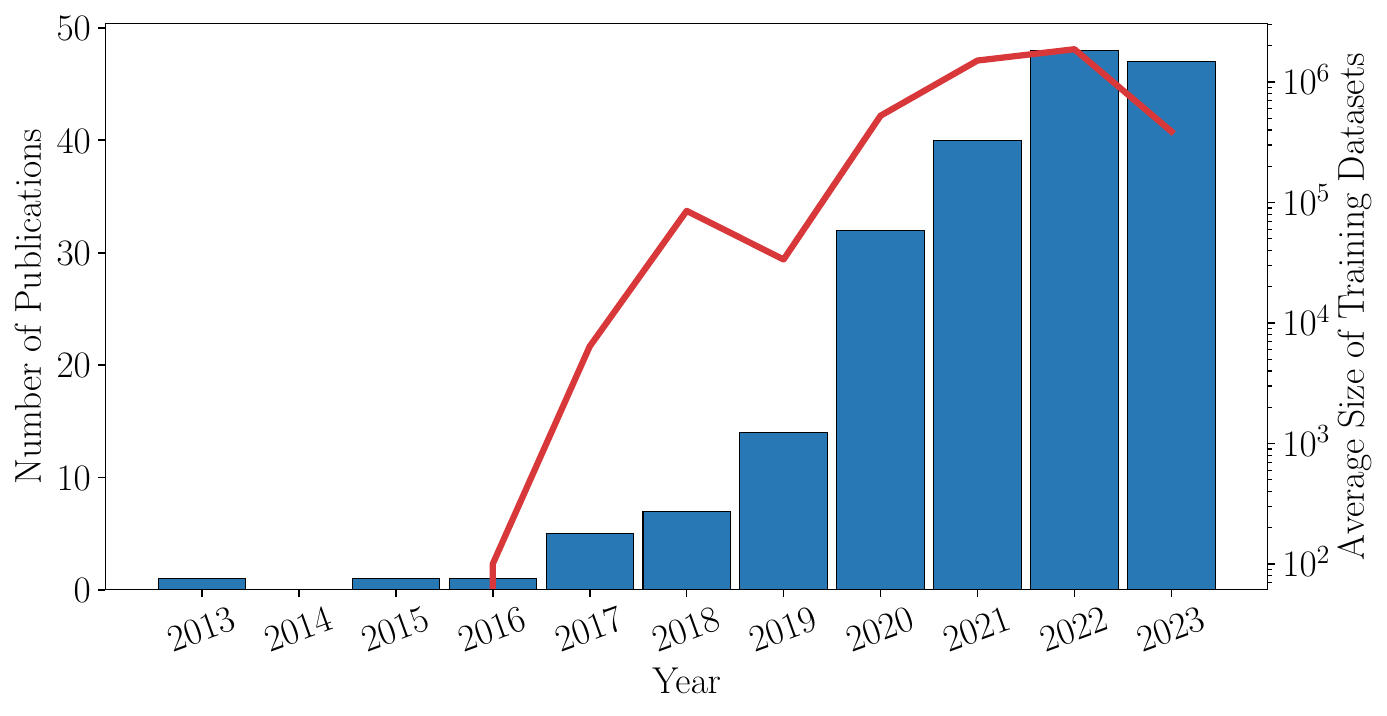}
	\caption{\textbf{Evolution of deep learning in gravitational wave data analysis (GWDA).} The green bars represent the number of published papers on GWDA employing deep learning techniques from 2013 to 2023. The red line traces the growth in the average size of training datasets used in these studies, highlighting the increasing reliance on larger datasets for enhanced model performance and generalization.}
	\label{fig:3}
\end{figure*}

\Ac{GWDA} is confronted with challenges stemming from the high-dimensional parameter space and the presence of non-Gaussian, non-stationary artifacts in the interferometer background \cite{covas_identification_2018}. The introduction of \ac{AI}, particularly \ac{DL} algorithms, offers a promising avenue to address these challenges \cite{bambi_advances_2021,cuoco_machine_2020}. These algorithms are characterized by their computational efficiency, leveraging accelerated hardware for rapid solutions \cite{benedetto_ai_2023}. Their scalability ensures they can handle extensive datasets, providing reliable model performance estimates \cite{huerta_accelerated_2021}. The modular nature of these algorithms facilitates adaptability, allowing for the streamlined incorporation of new methodologies \cite{ma_ensemble_2022}. Furthermore, the generalization capabilities of \ac{DL} models ensure consistent performance across varied \ac{GWDA} scenarios \cite{mcginn_generalised_2021}. Several recent reviews have explored the application of AI in \ac{GWDA}. Notably, Refs. \cite{benedetto_ai_2023,bambi_advances_2021,cuoco_machine_2020} focus primarily on ground-based, detectors, while Ref. \cite{cuoco_applications_2024} offers a more comprehensive overview of the field without providing an overall statistical analysis of the literature. In light of these advancements, our review undertakes:
\begin{itemize}
	\item The paper reviews various AI-driven approaches to \ac{GWDA}, \textbf{covering every stage} of the entire compact binary coalescences search pipeline, from waveform modeling to scientific discoveries.
	\item We offer the review of AI's application in gravitational
	      waveform modeling, showcasing novel methodologies and their accuracy.
	\item We also present the review of AI's role in dramatically
	      accelerating parameter estimation within gravitational wave studies.
	\item This paper presents \textbf{the most comprehensive review to date}, seamlessly
	      integrating the latest advancements in \ac{AI} with their practical
	      applications in the field of \ac{GWDA}.
	\item Our study culminates in a \textbf{meta-analysis}, synthesizing insights and
	      trends from diverse AI applications in \ac{GWDA},
	      discussing the fusion of AI and \ac{GWDA}, and opening
	      new avenues for insights and \textbf{future research directions}.
\end{itemize}
With this exploration, we aim to offer a panoramic view, intertwining the intricacies of \ac{AI} with the mysteries of \acp{GW}, underscoring the transformative potential of their collaboration.

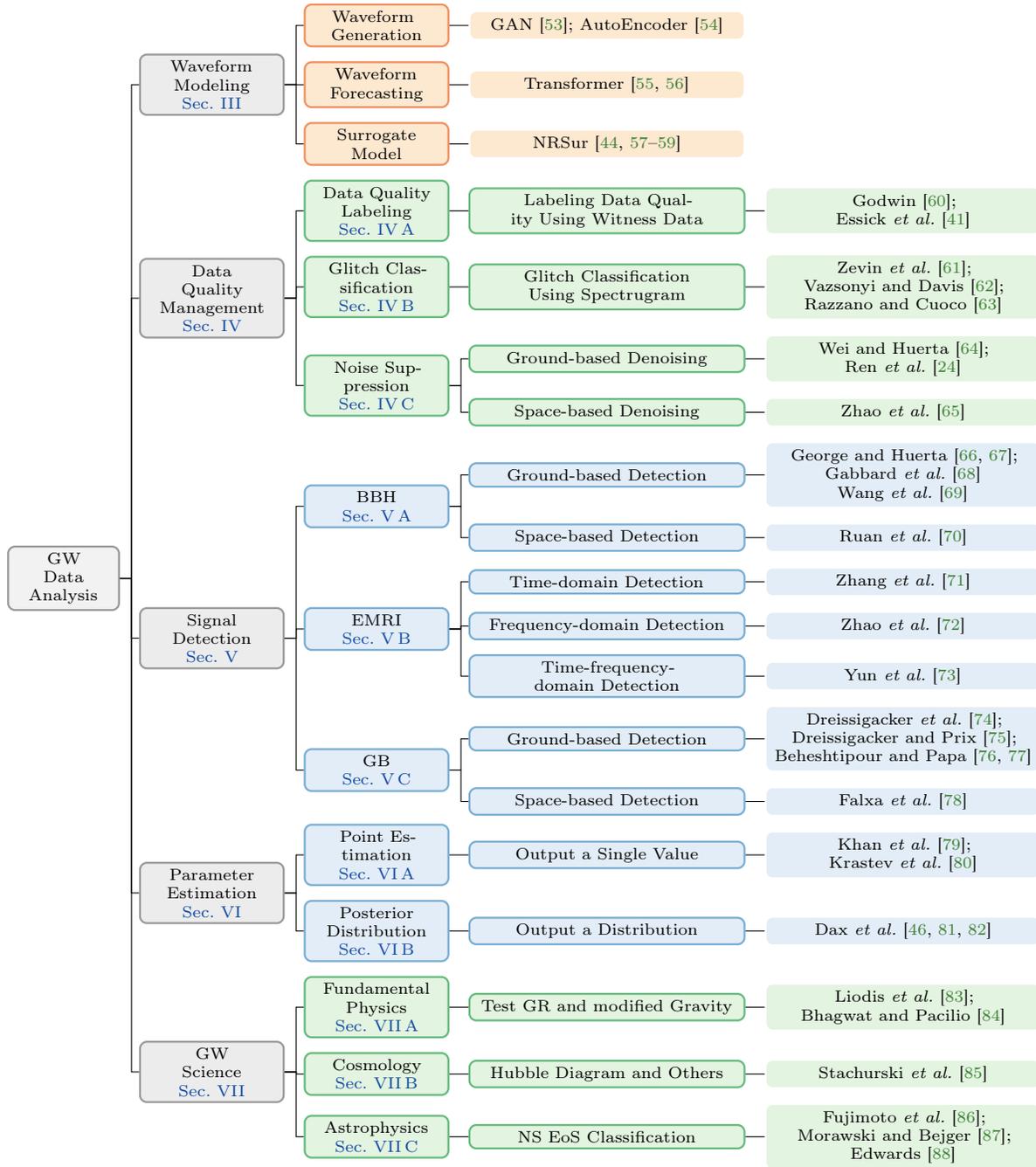
\begin{figure*}%
	\scriptsize
	\hspace*{-30pt}
	\begin{forest}
		for tree={
		forked edges,
		grow'=0,
		draw,
		rounded corners,
		node options={align=center,},
		text width=2.7cm,
		s sep=6pt,
		calign=edge midpoint,
		},
		[
		GW~\\ Data~\\ Analysis,
		fill=gray!45,
		parent
		[
		Waveform Modeling~\\ \Cref{sec:waveform},
		for tree={ top_class}
			[
				Waveform ~\\ Generation,
				for tree={fill=red!45,encoder}
					[
						GAN~\cite{mcginn_generalised_2021};
						AutoEncoder~\cite{liao_deep_2021},
						encoder_work
					]
			]
			[
				Waveform~\\ Forecasting,
				for tree={fill=red!45,encoder}
					[
						Transformer~\cite{khan_interpretable_2022-1,shi_compact_2023}\\,
						encoder_work
					]
			]
			[
				Surrogate Model,
				for tree={fill=red!45,encoder}
					[
						NRSur~\cite{varma_high-accuracy_2019,varma_surrogate_2019,varma_surrogate_2019-1,islam_surrogate_2022},
						encoder_work
					]
			]
		]
		[
		Data~\\ Quality~ \\ Management \\ \Cref{sec:dq}, for tree={ top_class}
		[
		Data Quality Labeling \\ \Cref{sec:dq_label},
		for tree={fill=green!45,generation}
		[
		Labeling Data Quality Using Witness Data,
		generation_more
		[
		\citet{godwin_low-latency_2020};\\
		\citet{essick_idq_2020},
		generation_work
		]
		]
		]
		[
		Glitch Classification \\ \Cref{sec:dq_glitch},
		for tree={fill=green!45,generation}
		[
		Glitch Classification Using Spectrugram,
		generation_more
		[
		\citet{zevin_gravity_2017};\\
		\citet{vazsonyi_identifying_2023};\\
		\citet{razzano_image-based_2018}\\,
		generation_work
		]
		]
		]
		[
		Noise Suppression \\ \Cref{sec:dq_denoise},
		for tree={fill=green!45,generation}
		[
		Ground-based Denoising,
		generation_more
		[
		\citet{wei_gravitational_2020-2};\\
		\citet{wang_waveformer_2024},
		generation_work
		]
		]
		[
		Space-based Denoising,
		generation_more
		[
		\citet{zhao_space-based_2023},
		generation_work
		]
		]
		]
		]
		[
		Signal~\\ Detection \\ \Cref{sec:detection},
		for tree={ top_class}
		[
		BBH \\ \Cref{sec:detection_bbh},
		for tree={fill=green!45,gpa}
		[
		Ground-based Detection,
		gpa_wide
		[
		\citet{george_deep_2018-4,george_deep_2018-5};\\
		\citet{gabbard_matching_2018}\\
		\citet{wang_gravitational_2020},
		gpa_work
		]
		]
		[
		Space-based Detection,
		gpa_wide
		[
		\citet{ruan_rapid_2023-2},
		gpa_work
		]
		]
		]
		[
		EMRI\\ \Cref{sec:detection_emri},
		for tree={fill=green!45,gpa}
		[
		Time-domain Detection,
		gpa_wide
		[
		\citet{zhang_detecting_2022},
		gpa_work
		]
		]
		[
		Frequency-domain Detection,
		gpa_wide
		[
		\citet{zhao_dilated_2024},
		gpa_work
		]
		]
		[
		Time-frequency-domain Detection,
		gpa_wide
		[
		\citet{Yun_detecting_2023},
		gpa_work
		]
		]
		]
		[
		GB \\ \Cref{sec:detection_gb},
		for tree={fill=green!45,gpa}
		[
		Ground-based Detection,
		gpa_wide
		[
		\citet{dreissigacker_deep-learning_2019-1};\\
		\citet{dreissigacker_deep-learning_2020};\\
		\citet{beheshtipour_deep_2020,beheshtipour_deep_2021},
		gpa_work
		]
		]
		[
		Space-based Detection,
		gpa_wide
		[
		\citet{falxa_adaptive_2023},
		gpa_work
		]
		]
		]
		]
		[
		Parameter~\\ Estimation \\ \Cref{sec:pe},
		for tree={ top_class}
		[
		Point Estimation \\ \Cref{sec:pe_point},
		for tree={fill=green!45,gpa}
		[
		Output a Single Value,
		gpa_wide
		[
		\citet{khan_physics-inspired_2020};\\
		\citet{krastev_detection_2021},
		gpa_work
		]
		]
		]
		[
		Posterior Distribution \\ \Cref{sec:pe_dist},
		for tree={fill=green!45,gpa}
		[
		Output a Distribution,
		gpa_wide
		[
		\citet{dax_real-time_2021,dax2022group,dax_neural_2023},
		gpa_work
		]
		]
		]
		]
		[
		GW~\\ Science \\ \Cref{sec:science},
		for tree={ top_class}
		[
		Fundamental Physics \\ \Cref{sec:science_phys},
		for tree={fill=green!45,generation}
		[
		Test GR and modified Gravity,
		generation_more
		[
		\citet{liodis_neural-network-based_2023};\\
		\citet{bhagwat_merger-ringdown_2021},
		generation_work
		]
		]
		]
		[
		Cosmology\\ \Cref{sec:science_cosmos},
		for tree={fill=green!45,generation}
		[
		Hubble Diagram and Others,
		generation_more
		[
		\citet{stachurski_cosmological_2023},
		generation_work
		]
		]
		]
		[
		Astrophysics \\ \Cref{sec:science_astroph},
		for tree={fill=green!45,generation}
		[
		NS EoS Classification,
		generation_more
		[
		\citet{fujimoto_mapping_2020};\\
		\citet{morawski_neural_2020};\\
		\citet{edwards_classifying_2021}
		, generation_work
		]
		]
		]
		]
		]
	\end{forest}
	\caption{\textbf{The structure of our review.} This tree chart delineates the structured progression of our in-depth exploration into the convergence of gravitational waves and artificial intelligence. Each branch represents a dedicated subdomain, emphasizing the harmonious integration of gravitational wave data analysis with the innovative strides of deep learning techniques.}
	\label{fig:paper_structure}
\end{figure*}

The remainder of this review is structured to provide a comprehensive overview
of the intersection between \ac{AI} and \acp{GWDA}
(\Cref{fig:paper_structure}). In \Cref{sec:dl}, we briefly introduce the
development of \ac{DL}. In \Cref{sec:waveform}, we delve into waveform
modeling, discussing both traditional methods and the advancements brought
about by \ac{DL}. \Cref{sec:dq} is dedicated to data quality management,
encompassing topics from data quality labeling to noise suppression. Signal
detection, a pivotal stage in the analysis, is covered in \Cref{sec:detection},
where we explore various methods and their implications. Parameter estimation,
a complex yet crucial aspect, is dissected in \Cref{sec:pe}. Moving beyond the
technicalities, \Cref{sec:science} sheds light on the broader scientific
discoveries enabled by these methodologies. Lastly, in \Cref{sec:discussion},
we offer a meta-analysis of the literature, discuss futuristic insights, and
chart potential directions for the field.

\begin{figure*}
	\centering
	\includegraphics[width=0.6\textwidth]{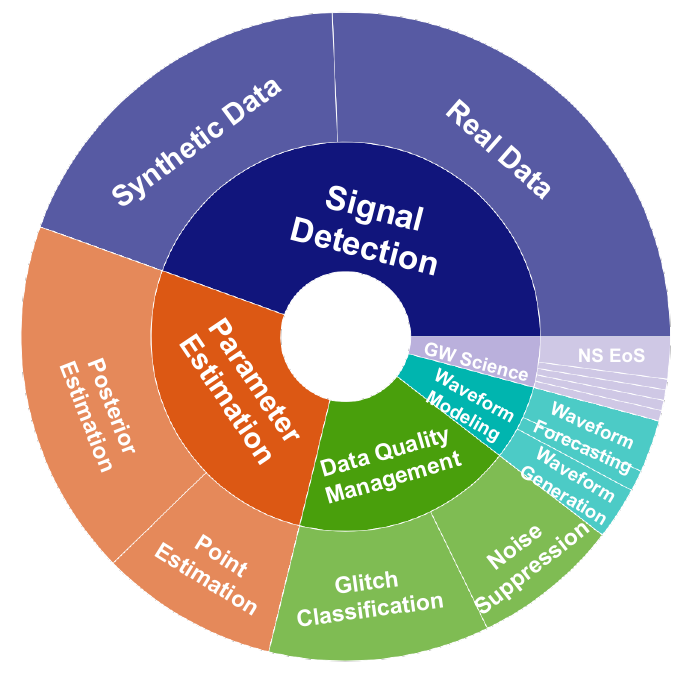}
	\caption{\textbf{Hierarchical breakdown of deep learning applications in gravitational wave data analysis.}  The pie chart provides a visual distribution of published papers on GWDA using deep learning, categorized by specific subdomains. Each segment represents a distinct area of application, showcasing the diversity and breadth of deep learning techniques in advancing gravitational wave research.}
	\label{fig:4}
\end{figure*}

\section{Deep Learning}
\label{sec:dl}

% ----------------------
% --------------------------

With AlexNet's unprecedented performance in the ImageNet challenge, 2012 marked
an important turning point in \ac{AI} \cite{krizhevsky_imagenet_2012}. This
accomplishment demonstrated the potential of neural networks for complex visual
tasks. \Acp{CNN} rapidly ascended as the benchmark in image processing,
leveraging spatial hierarchies through convolutional layers and pooling
operations \cite{simonyan_very_2015}. However, as network architectures
deepened, the vanishing gradient problem became evident. \Ac{ResNet} addressed
this, introducing skip connections to facilitate gradient flow
\cite{he_deep_2016}. Concurrently, for sequential data, \acp{RNN} emerged as a
promising approach \cite{cho_learning_2014}. Their capacity to retain state was
notable, but challenges with long-term dependencies persisted. WaveNet
\cite{oord_wavenet_2016}, designed for raw audio generation, addressed this by
employing dilated convolutions, expanding the receptive field. In the
unsupervised learning domain, AutoEncoders \cite{hinton_reducing_2006}, with
their encoder-decoder structure, excelled at deriving efficient data
representations without relying on labels. Together, these innovations
post-AlexNet have shaped the trajectory of \ac{DL}, setting the stage for
subsequent advancements.

Contrasted with discriminative models that have been the mainstay of \ac{DL},
generative modeling has carved its own niche, offering unique capabilities
\cite{ganguli_predictability_2022}. \Acp{GAN} \cite{goodfellow_2014} emerged as
a groundbreaking approach, where two neural networks, a generator and a
discriminator, engage in a game-theoretic framework \cite{Bowles_gan_2018}. The
generator crafts synthetic data, while the discriminator discerns between real
and generated samples. This adversarial process results in the generator
producing increasingly realistic data \cite{shmelkov_how_2018}. \Acp{VAE}
\cite{kingma_auto-encoding_2014} offered another perspective, framing
generative modeling as a probabilistic graphical model where the encoder and
decoder networks are conditioned on latent variables. Normalizing flows
\cite{rezende_variational_2015} and Diffusion models \cite{ho_denoising_2020}
further enriched the generative landscape, providing mechanisms to transform
simple distributions into complex data distributions. These generative models
have been instrumental in the rise of \ac{AIGC} \cite{cao_comprehensive_2023},
enabling the synthesis of high-fidelity images \cite{xiang_gram-hd_2023},
videos \cite{gao_high-fidelity_2023}, and even art
\cite{oppenlaender_creativity_2022}. The ability of these models to generate
content that is often indistinguishable from real-world data has opened new
avenues in digital media \cite{karnouskos_artificial_2020}, virtual reality
\cite{geraci_apocalyptic_2010}, and beyond.

As the field of \ac{DL} progressed, the Transformer architecture emerged as a
groundbreaking innovation, particularly for sequence-based tasks. Introduced by
\citet{vaswani_attention_2017}, the Transformer discarded the recurrent layers
that characterized \acp{RNN}, instead relying on self-attention mechanisms to
process input data. This self-attention allows the model to weight the
significance of different parts of an input sequence, regardless of their
positional distance, enabling it to capture long-range dependencies in the data
with ease \cite{choromanski_rethinking_2023}. Unlike \acp{CNN}, which uses
fixed-size filters to process data in local receptive fields, the Transformer's
self-attention mechanism provides it with a dynamic receptive field, adjusting
based on the content of the input \cite{brasoveanu_visualizing_2020}. This
flexibility allows Transformers to excel in tasks where the importance of data
points varies contextually \cite{garg_what_2022}. Furthermore, while \acp{RNN}
processes sequences step-by-step, leading to potential issues with long
sequences due to vanishing and exploding gradients, transformers process all
sequence elements in parallel. This parallel processing not only speeds up
training but also alleviates the gradient issues associated with long sequences
\cite{dao_flashattention-2_2023}. Building on the foundational transformer
architecture, \acp{LLM} like BERT \cite{devlin_bert_2019} and GPT
\cite{Radford_Language_2019} have set new benchmarks in natural language
processing. ChatGPT, encompassing iterations like GPT-3.5 and GPT-4
\cite{openai_gpt-4_2023}, epitomizes the rapid evolution and capabilities of
\acp{LLM}. These models, with their ability to understand, generate, and reason
with text, have showcased the superiority of the transformer architecture over
traditional \acp{CNN} and \acp{RNN} in handling sequence data
\cite{karita_comparative_2019}.

While supervised learning and generative models have made significant strides, another paradigm, \ac{RLHF}, has emerged as a potent tool in the \ac{AI} arsenal \cite{ramamurthy_is_2022}. At its core, reinforcement learning is an approach where an agent learns to maximize rewards by interacting with an environment, much like how one might learn a game through trial and error \cite{mataric_reward_1994}. However, designing suitable reward functions for complex tasks can be challenging. \ac{RLHF},sidesteps these challenges
by leveraging human feedback as a primary source of reward signals
\cite{griffith_policy_2013}. This approach allows models to learn more complex
behaviors without the need for explicit reward shaping. OpenAI's fine-tuning of
models like GPT-4 using \ac{RLHF} is a testament to the power of this approach
\cite{openai_gpt-4_2023}. By collecting comparison data, ranking different
model responses, and using proximal policy optimization, models can be
fine-tuned to produce safer and more useful outputs \cite{gu_proximal_2022}.
Simultaneously, the field of multi-agent systems is undergoing substantial
development \cite{vamvoudakis_multi-agent_2021}. As \ac{AI} models become more
sophisticated, there's a growing interest in how they interact in shared
environments \cite{christianos_shared_2020}. Multi-agent systems, where
multiple \ac{AI} entities collaborate or compete, offer insights into emergent
behaviors, cooperation strategies, and even the evolution of communication
\cite{nguyen_deep_2020}. The fusion of \acp{LLM} with multi-agent setups is
particularly intriguing \cite{chan_chateval_2023,wu_autogen_2023}. Imagine
agents equipped with the linguistic prowess of \acp{LLM} negotiating,
strategizing, and evolving their communication protocols in real-time. Such
advancements hint at a future where \ac{AI} entities don't just operate in
isolation but actively learn from, compete with, and collaborate with other
\ac{AI} entities, paving the way for more dynamic and adaptive \ac{AI}
ecosystems \cite{talebirad_multi-agent_2023}.

\section{Waveform Modeling}
\label{sec:waveform}

\ac{GW} astronomy has experienced significant progress in recent years, with the numerical modeling of compact binary coalescence—including BBH, BNS, and NSBH—which has been a key component in developing accurate waveform templates \cite{pretorius_evolution_2005}. The journey of these cosmic phenomena is typically segmented into three phases: the inspiral, merger, and ringdown, with the potential for tidal disruptions in systems involving neutron stars \cite{shibata_simulation_2000}. Moreover, burst signals arising from events such as core-collapse supernovae or tidal disruption events contribute another class of short-duration waveforms, which, despite their less predictable morphology, are crucial for a complete understanding of GW sources. The accurate depiction of these stages through gravitational waveforms is vital for the detection of \ac{GW} signals and the interpretation of the celestial narratives they unveil \cite{baker_gravitational-wave_2006,campanelli_accurate_2006}.

\begin{table*}[htbp]
	\begin{center}
		\caption{\textbf{AI-Driven Waveform Modeling in Gravitational Wave Astronomy.} This table illustrates the accuracy of typical AI models in generating GW waveforms. The performance of each model is evaluated by the overlap \cite{purrer_gravitational_2020}, which quantifies the similarity between AI-generated waveforms and standard templates.}
		\label{tab:waveform_modeling}%
		\begin{tabular}{@{}llcc@{}}
			\toprule
			\hline
			\textbf{Paper}                    & \textbf{Task}        & \textbf{Model} & \textbf{Overlap} \\
			\hline\hline
			\citet{huerta_eccentric_2018}     & Waveform Generation  & GP             & 0.99             \\
			\citet{mcginn_generalised_2021}   & Waveform Generation  & GAN            & --               \\
			\citet{liao_deep_2021};           & Waveform Generation  & CVAE           & 0.9841           \\
			\citet{khan_interpretable_2022-1} & Waveform Forecasting & Transformer    & 0.993            \\
			\citet{islam_surrogate_2022}      & Surrogate Modeling   & MLP            & 0.99             \\
			\hline
			\bottomrule
		\end{tabular}
	\end{center}
\end{table*}

The creation of waveform templates is an essential aspect of \ac{GW} data
analysis. These templates act as blueprints for expected \ac{GW} signals from
diverse astrophysical origins and are crucial for the comparison with data
collected by \ac{GW} observatories. However, the production of these templates,
particularly for intricate systems, poses significant computational challenges.
The computational demands and the sparse nature of \ac{NR} waveforms
necessitate the use of more practical approaches. To this end, a variety of
approximate waveform models have been developed, such as the \texttt{IMRPhenom}
\cite{khan_frequency-domain_2016,pratten_computationally_2021,yu_imrphenomxode_2023}
and \texttt{SEOBNR}
\cite{han_constructing_2011,cao_waveform_2017,mihaylov_pyseobnr_2023} families,
which are calibrated against \ac{NR} simulations to ensure accuracy and
efficiency in \ac{GW} data analysis.

In parallel with these developments, the field of waveform modeling is
undergoing a transformative phase with the introduction of \ac{DL} techniques.
\ac{DL} offers a promising pathway to expedite the generation of waveform
templates by employing neural networks that are potentially capable of quickly synthesizing the complex dynamics of compact binary mergers. This innovative approach could speed up the process of generating waveforms,potentially overcoming some of the computational bottlenecks of traditional methods. For a detailed overview of the advancements in waveform modeling facilitated by DL, refer to \Cref{tab:waveform_modeling}.

Although the evolution of \ac{GW} waveforms is fundamentally deterministic and governed by differential equations, forecasting in the deep learning context refers to training a model to implicitly learn the waveform evolution from data \cite{fons_hypertime_2022}. This data-driven approach enables rapid predictions that are useful for real-time analysis \cite{khan_interpretable_2022-1}. \ac{DL} models, particularly transformers, have the potential to revolutionize waveform
forecasting. \citet{khan_interpretable_2022-1} stands as a testament to this
potential, showcasing the capabilities of \ac{DL} in predicting the evolution
of \ac{GW} signals.

In \ac{GW} astronomy, the imperative for swift and accurate waveform modeling
is ever-present \cite{gamboa_accurate_2024}. \ac{DL} presents a promising pathway, but the intricate
parameter space of \ac{GW} signals can be daunting \cite{khan_gravitational-wave_2021}. Surrogate models, trained
on \ac{NR} waveforms, integrate \ac{DL} techniques to expedite
traditional waveform calculations \cite{islam_surrogate_2022}. These models
stand as rapid waveform generators, adeptly spanning the vast \ac{GW} parameter
space \cite{huerta_eccentric_2018}. Typically, the speed-up factor achieved through surrogate models is greater than 100, making them significantly faster than traditional methods \cite{shi_rapid_2025}.Recent investigations indicate that surrogate models may substantially enhance computational efficiency, offering a promising approach to align traditional \ac{GW} methodologies with the capabilities of AI \cite{freitas_nrsurnn3dq4_2024}.

\section{Data Quality Management}
\label{sec:dq}

Ensuring high data quality is essential in GW astronomy, where detecting these subtle spacetime ripples depends on precise measurements. This section examines the critical aspects of data quality management, including accurate data labeling, robust glitch classification, and effective denoising. Each of these procedures is vital for maintaining data integrity and reliability, which are fundamental to successful GW detection.

\subsection{Data Quality Labeling}
\label{sec:dq_label}

The detection of \ac{GW} is a complex task, primarily due to the challenge of
distinguishing these faint cosmic signals from background noise. The
sensitivity of \ac{GW} detectors is such that they pick up a myriad of noises,
ranging from seismic activities to instrumental glitches
\cite{davis_detector_2022}. This noise can often mimic or obscure the actual
\ac{GW} signals, making the task of identifying genuine events extremely
challenging \cite{pankow_mitigation_2018,coughlin_classifying_2019}. The
intricate nature of \ac{GW} signals, often buried deep within the noise,
requires sophisticated methods to accurately separate signals from noise. This
complexity is a fundamental difficulty in \ac{GW} detection, necessitating
advanced techniques for noise analysis and signal extraction
\cite{wang_waveformer_2024}.

The quality of the data plays a crucial role in the successful detection and
analysis of \ac{GW} signals. High-quality data ensures that the signals
extracted are accurate representations of astrophysical events. To achieve
this, it is essential to label and categorize the data effectively
\cite{davis_2021}. Traditional methods of data quality monitoring involve
manual labeling, which, while thorough, can be time-consuming and subject to
human error \cite{bekker_training_2016}. The vast volume of data generated by
\ac{GW} detectors adds to the complexity, as each data segment needs to be
meticulously analyzed to determine its quality. This process is vital as it
directly impacts the reliability of signal detection and the subsequent
astrophysical interpretations \cite{davis_2021}.

\ac{DL} has emerged as a powerful tool in addressing the challenges of data quality labeling in \ac{GW} astronomy \cite{godwin_low-latency_2020,razzano_image-based_2018}. \ac{DL} algorithms, particularly neural networks, have the capability to automate the process of data quality assessment, offering both efficiency and accuracy \cite{glanzer_data_2023}. These algorithms can process large volumes of data rapidly, identifying patterns and anomalies that may indicate issues with data quality \cite{biswas_application_2013}. Several studies have demonstrated the effectiveness of \ac{DL} in automating the data quality labeling process, significantly reducing the time and effort required while maintaining, if not improving, the accuracy of the labels \cite{essick_idq_2020,fernandes_convolutional_2023,cuoco_applications_2024}. By ensuring that only high-quality data is fed into analysis pipelines, \ac{DL} contributes significantly to the robustness and reliability of \ac{GW} signal detection and analysis \cite{davis_2021}.

\subsection{Glitch Classification}
\label{sec:dq_glitch}

\begin{figure}
	\centering
	\includegraphics[width=0.45\textwidth]{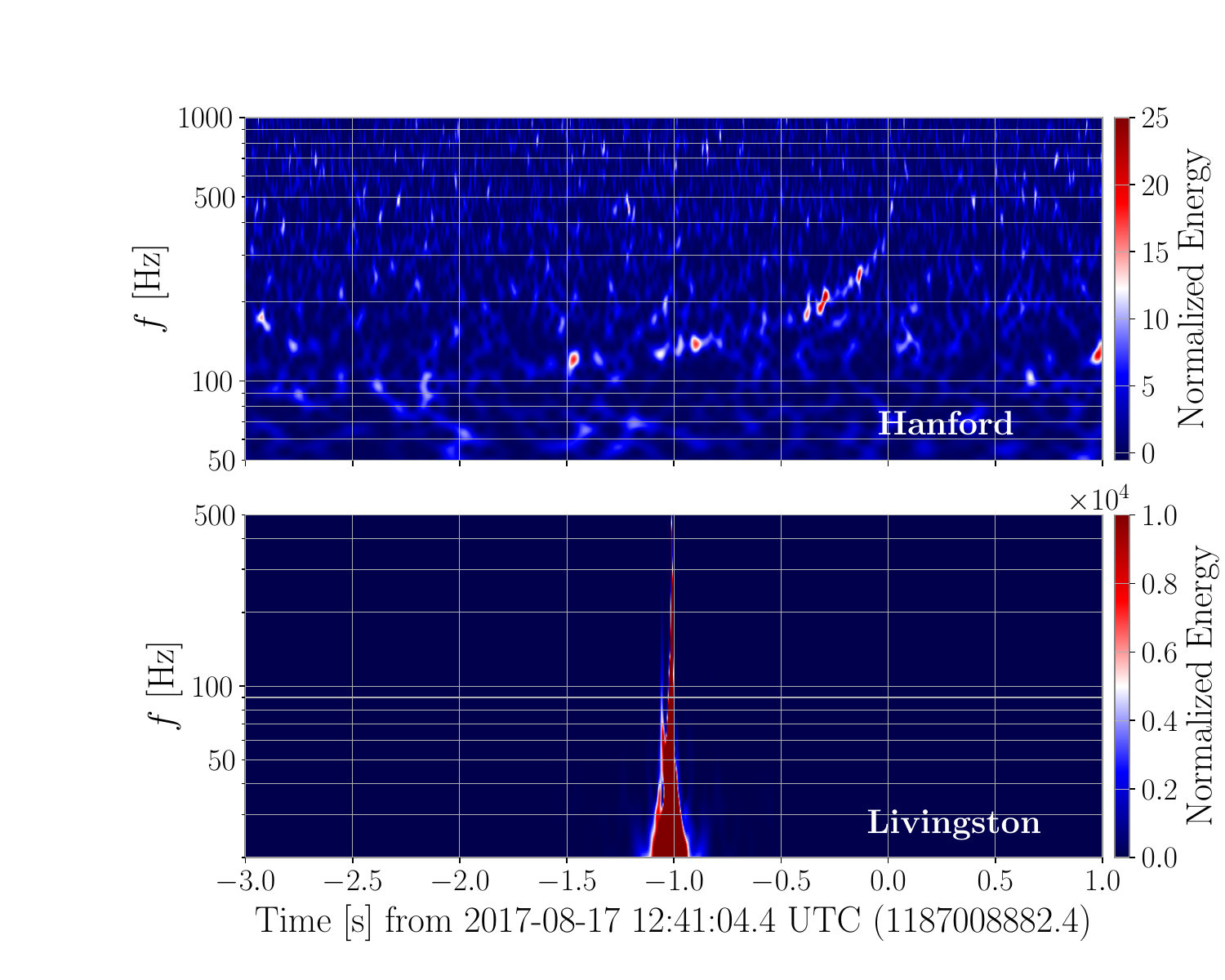}
	\caption{\textbf{The influence of glitches on GW detection.} This figure presents raw data from the LIGO Hanford and Livingston observatories, capturing the moment of GW170817. It vividly illustrates how glitches, with their power surpassing that of the actual signal, can significantly obstruct gravitational wave detection.}
	\label{fig:glitch}
\end{figure}

Glitches, or transient noise artifacts, are a recurring challenge in \ac{GW}
data analysis. These anomalies can significantly impede signal detection
processes (see \Cref{fig:glitch}), necessitating their accurate classification
and understanding \cite{merritt_transient_2021}. The impact of glitches is
twofold: they can either mimic authentic GW signals, leading to false
positives, or they can obscure genuine signals, resulting in missed detections
\cite{nitz_distinguishing_2018}. This dual nature of glitches makes them a
critical focus in GW data analysis, as their presence can skew the results and
interpretations of GW observations.

Traditionally, the classification of glitches has relied on a variety of
methods. Notable initiatives in this realm are the Gravity Spy and GWitchHunters project, which
have been instrumental in tackling the glitch challenge
\cite{zevin_gravity_2017,razzano_gwitchhunters_2023}. This project has compiled a comprehensive dataset
that serves as a crucial resource for the classification of glitches.
Traditional methods often involve manual inspection or semi-automated
techniques, where analysts use witness data—auxiliary information from the
detectors—to help identify and categorize these glitches. These methods, while
effective, can be time-consuming and may not always capture the subtle nuances
of different glitch types \cite{mukherjee_classification_2010}.

\begin{table}[htbp]
	\begin{center}
		\caption{\textbf{Comparative analysis of AI techniques in glitch classification.} The table presents a comparison of various AI methodologies, all evaluated using the Gravity Spy dataset. Given the nature of this task as a classification challenge, accuracy serves as the metric for assessing the effectiveness of each approach.}
		\label{tab:glitch}%
		\begin{tabular}{@{}lcc@{}}
			\toprule
			\hline
			\textbf{Paper}                                                & \textbf{Model} & \textbf{Accuracy} \\
			\hline\hline
			\citet{mukund_transient_2017}                                 & DBNN           & 0.99              \\
			\citet{powell_classification_2015,powell_classification_2017} & WDF            & 0.92              \\
			\citet{powell_2023}                                           & GAN            & 0.99              \\
			\citet{razzano_image-based_2018}                              & SVM            & 0.971             \\
			\citet{razzano_image-based_2018}                              & CNN            & 0.998             \\
			\citet{bahaadini_machine_2018}                                & SVM            & 0.9821            \\
			\citet{soni_discovering_2021}                                 & CNN            & 0.988             \\
			\citet{sakai_2022a}                                           & CNN            & 0.97              \\
			\citet{fernandes_convolutional_2023}                          & ConvNeXt       & 0.981             \\
			\citet{george_classification_2018-2}                          & ResNet         & 0.988             \\
			\hline
			\bottomrule
		\end{tabular}
	\end{center}
\end{table}

The advent of \ac{DL} has introduced a new paradigm in glitch classification,
offering more efficient approaches \cite{fernandes_convolutional_2023}. Some of these techniques directly analyze time-series data from GW detectors, identifying patterns that are characteristic of specific types of glitches \cite{vajente_machine-learning_2020}. Another innovative approach involves transforming time-series data into visual formats, allowing \acp{CNN} to classify glitches based on their visual signatures
\cite{powell_classification_2015,powell_classification_2017,razzano_image-based_2018,fernandes_convolutional_2023}. This method leverages the pattern recognition
capabilities of \acp{CNN} to discern between different glitch types
effectively. Looking ahead, the integration of \ac{DL} in glitch classification
is poised to play a pivotal role in the future of \ac{GW} data analysis,
potentially leading to more accurate detections and a deeper understanding of
the cosmos \cite{razzano_gwitchhunters_2023}. For a detailed exploration of
these \ac{DL} methods, refer to \Cref{tab:glitch}.

\subsection{Data Denoising}
\label{sec:dq_denoise}

\begin{figure}
	\centering
	\includegraphics[width=0.45\textwidth]{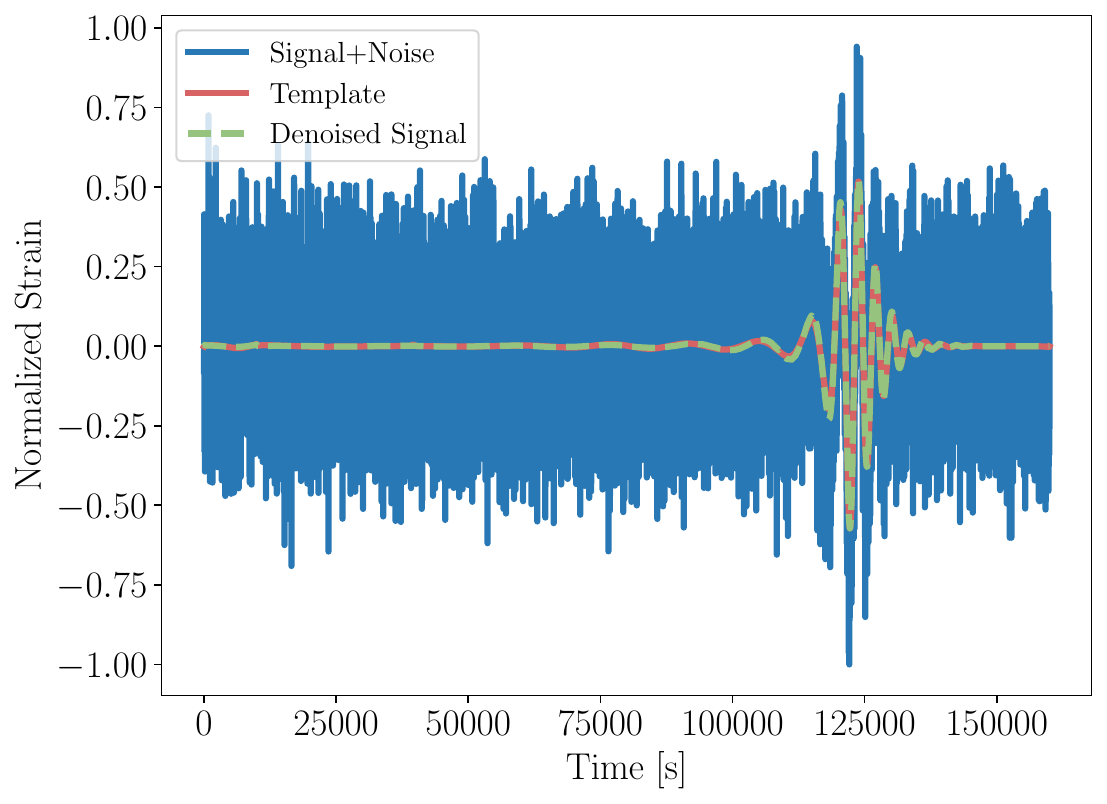}
	\caption{\textbf{GW denoising by the transformer-based model.}  The purple line depicts the original signal, buried in noise, while the green line shows the theoretical signal template. The orange dashed line represents the denoised signal, highlighting the model's efficacy in noise reduction. Adapted from Ref. \cite{zhao_space-based_2023}.}
	\label{fig:denoise}
\end{figure}

In the realm of \ac{GW} astronomy, noise is an unavoidable and challenging
factor that significantly impacts signal detection
\cite{nitz_distinguishing_2018}. The presence of noise in the data result in a low \ac{SNR}, leading to difficulties in identifying genuine
\ac{GW} events \cite{finn_detection_1992}. Effective denoising is therefore
crucial, as it not only improves the SNR but also decreases the \ac{FAR},
making it easier to distinguish real signals from noise. Additionally,
denoising plays a vital role in the convergence of \ac{MCMC} parameter
estimation processes. By suppressing the noise, denoising methods help in
faster and more accurate extraction of signal from the noisy data, thereby
enhancing the overall efficiency and reliability of GW signal analysis
\cite{wang_waveformer_2024}.

Traditional denoising methods in \ac{GW} astronomy have focused on direct noise
suppression techniques, encompassing a variety of approaches \cite{torres-forne_total-variation_2018,cornish_bayeswave_2015,klimenko_coherent_2008}. These include variational-based methods \cite{torres-forne_total-variation_2018}, wavelet-based methods, and techniques utilizing the Hilbert-Huang transform \cite{akhshi_template-free_2021}. Each of these strategies aims to remove noise components while preserving the integrity of potential \ac{GW} signals. For instance, variational-based methods apply mathematical optimization techniques to filter out noise \cite{torres_total-variation-based_2014}, while wavelet-based methods use wavelet transformations to reconstruct signals from data \cite{cornish_bayeswave_2021,klimenko_method_2016}. The Hilbert-Huang transform, a novel approach, is particularly effective for non-linear and non-stationary data \cite{akhshi_template-free_2021}. While effective in certain scenarios, these traditional approaches can be limited in their ability to handle the complex and dynamic nature of noise in \ac{GW} data without the help of data from witness channel \cite{abbott_guide_2020}.

Although DL-based denoising methods are not yet integrated into current detection pipelines, research has shown that they have advanced signal extraction techniques in GWDA. These include dictionary learning
\cite{torres-forne_denoising_2016}, which employs basis waveforms for signal
representation; WaveNet \cite{wei_gravitational_2020} for its sequential data
handling capabilities; denoising autoencoders \cite{shen_denoising_2019} for
reconstructing signals from noisy data; and \ac{RNN}
\cite{chatterjee_extraction_2021} for capturing temporal dependencies.
Notably, the transformer-based model, as discussed in Ref. \cite{wang_waveformer_2024,zhao_space-based_2023}, has been applied to both ground-based and space-based \ac{GW} denoising and detection (\Cref{fig:denoise}), showcasing its potential in handling complex \ac{GW} data, the transformer-based approach demonstrated measurable improvements in sensitivity and a reduction in false alarm rates compared to traditional methods, thereby underscoring its practical utility in real-world applications.

\section{Signal Detection}
\label{sec:detection}
\subsection{BBH}
\label{sec:detection_bbh}

The detection of \acp{GW}, is a formidable task,
primarily due to the subtlety of these signals against a backdrop of
overwhelming instrumental and terrestrial noise. The advent of interferometric detectors, notably LIGO and Virgo, has revolutionized GW astronomy by enabling the first direct detection of GW event. However, the increased
sensitivity and observational bandwidth of these instruments have also
introduced intricate data analysis challenges. A cornerstone of \ac{GWDA} has
been matched filtering \cite{finn_detection_1992}, a technique that
cross-correlates observed data with theoretical waveform templates. Given the
high-dimensional parameter space associated with potential astrophysical
sources, this method, while powerful, is computationally demanding, especially
in the context of real-time analysis.

\begin{figure}
	\centering
	\includegraphics[width=0.45\textwidth]{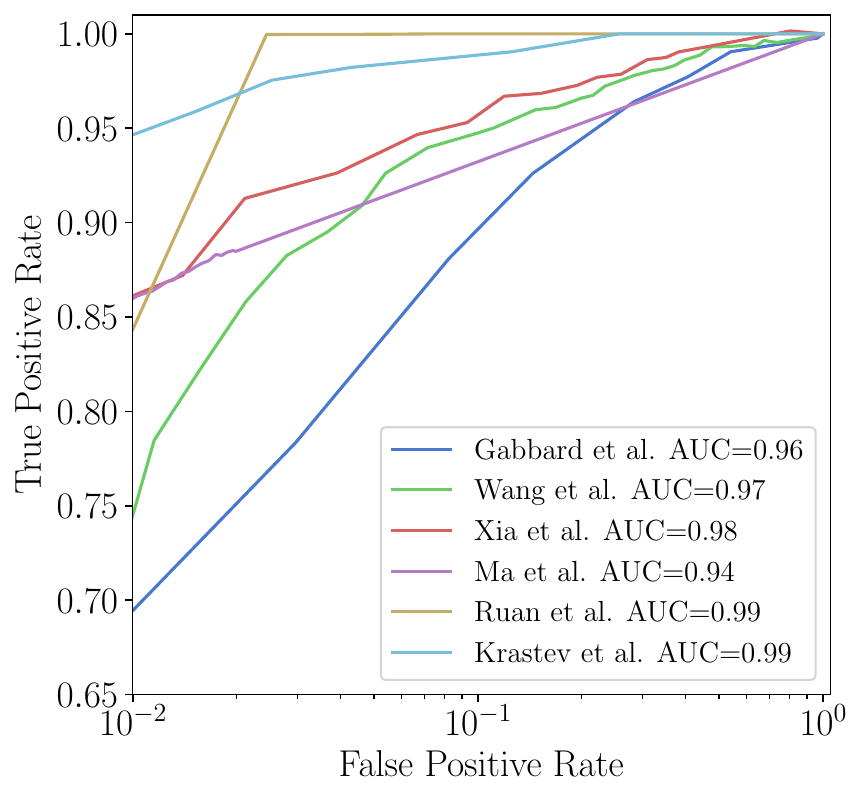}
	\caption{\textbf{Performance characterization of models' detection performance on synthetic data.} This figure presents the receiver operating characteristic (ROC) curves of various deep learning models, as detailed in \Cref{tab:detection_mock}. These curves illustrate the performance of each classifier in GW detection on synthetic data, effectively capturing their true positive rate against the false positive rate, thereby providing a clear visual representation of their efficacy in distinguishing between signal and noise. All data points are extracted from the original papers.}
	\label{fig:roc}
\end{figure}

Machine learning, particularly deep learning, has emerged as a potential tool for addressing the computational challenges in GW detection \cite{george_deep_2018-4,george_deep_2018-5}. The initial forays into integrating \ac{DL} algorithms with mock \ac{LIGO} data
underscored the viability and promise of these approaches
\cite{george_deep_2018-4,george_deep_2018-5,jiang_identify_2022,li_optimizations_2020,luo_extraction_2020}.
These models, trained on simulated waveforms, were subsequently tested on
synthetic data. Impressively, their performance in discerning potential \ac{GW}
signatures from noise was on par with traditional matched filtering-based
techniques, highlighting an important step forward in \ac{GWDA}
\cite{gabbard_matching_2018}, \Cref{tab:detection_mock} lists out some related
works, and their \ac{AUC} for detailed performance comparison corresponding
\ac{ROC} curve is depicted in \Cref{fig:roc}.

\begin{table}[htbp]
	\begin{center}
		\caption{\textbf{Comparative evaluation of AI methods for BBH GW detection on synthetic data.} This table showcases a detailed comparison of different AI techniques, each tested on synthetic data for BBH GW detection. As this task involves binary classification, the Area Under the Curve (AUC) is utilized as the key metric to gauge the performance and accuracy of these AI models.}
		\label{tab:detection_mock}%
		\begin{tabular}{@{}llc@{}}
			\toprule
			\hline
			\textbf{Paper}                   & \textbf{Model} & \textbf{AUC} \\
			\hline\hline
			\citet{gabbard_matching_2018};   & CNN            & 0.96         \\
			\citet{wang_gravitational_2020}  & MFCNN          & 0.97         \\
			\citet{xia_improved_2021}        & CNN            & 0.98         \\
			\citet{ma_ensemble_2022}         & CNN            & 0.94         \\
			\citet{ruan_rapid_2023-2}        & MFCNN          & 0.99         \\
			\citet{krastev_real-time_2020-2} & CNN            & 0.99         \\
			\hline
			\bottomrule
		\end{tabular}
	\end{center}
\end{table}

In light of the proven effectiveness of machine learning on simulated \ac{LIGO}
datasets, the subsequent logical step was its application to genuine
observational data. This transition presented inherent complexities, primarily
attributed to the non-Gaussian and non-stationary nature of the noise, which
elevated false alarm rates. Addressing these challenges,
\citet{wang_gravitational_2020} integrated a matched filtering sensing layer
prior to the \ac{CNN}, showcasing its performance on O1 data. Building upon
foundational research \cite{wei_gravitational_2020-2,wei_deep_2021-5},
\citet{huerta_accelerated_2021} employed a hardware-accelerated WaveNet to
probe for \ac{GW} signals within a month-long span of \ac{LIGO} data.
Concurrently, \citet{wang_waveformer_2024} proposes WaveFormer, a deep
learning-based data quality enhancement method for real observational
gravitational wave processing, achieving state-of-the-art noise suppression
performance, which paves a solid foundation for future strides in \ac{GW} data
processing and search \cite{skliris_real-time_2021,skliris_machine_2021}. See \Cref{tab:detection_real} for a detailed performance comparison of \ac{FAR}.

As the \ac{GW} community approaches a transformative phase, the ongoing development of space-based and third-generation ground-based detectors heralds a new observational epoch. These advanced instruments, set to provide
unparalleled observational capabilities, also bring forth intricate data
analysis challenges, particularly due to overlapping \ac{GW} signals
\cite{speri_roadmap_2022}. Recent endeavors on \ac{LISA} \cite{amaro-seoane_laser_2017} and \ac{ET} \cite{maggiore_science_2020} mock data
have shown promising results \cite{ruan_rapid_2023-2,zhao_space-based_2023,alhassan_detection_2023}. The ongoing advancement and incorporation of machine learning techniques remain crucial for effectively addressing these challenges in the forthcoming era.

\begin{table}[htbp]
	\begin{center}
		\caption{\textbf{Comparative analysis of AI approaches for BBH GW detection using detector strain data.} The table provides a comparison of various AI methodologies applied to the detection of GWs in detector strain data. In this context, where the focus is on minimizing erroneous detections, the false alarm rate (FAR) of GW signals is employed as the primary metric to evaluate the effectiveness and precision of each AI technique.}
		\label{tab:detection_real}%
		\begin{tabular}{@{}llc@{}}
			\toprule
			\hline
			\textbf{Paper}                     & \textbf{Model} & \textbf{FAR}                   \\
			\hline\hline
			\citet{krastev_detection_2021}     & CNN            & $\mathcal{O} (10^3)$ per month \\
			\citet{zhang_deep_2022}            & BiGRU          & 1 per 18.2 hours               \\
			\citet{wei_deep_2021-5};           & WaveNet        & 1 per 2.7 days                 \\
			\citet{tian_physics-inspired_2023} & GNN            & 1 per month                    \\
			\citet{schafer_one_2022}           & CNN            & 1 per $10^4$ months            \\
			\citet{wang_waveformer_2024}       & Transformer    & 1 per 1000 years               \\
			\hline
			\bottomrule
		\end{tabular}
	\end{center}
\end{table}

\subsection{EMRI}
\label{sec:detection_emri}
\Acp{EMRI} are among the most intriguing astrophysical phenomena in the universe. These events occur when a stellar-mass compact object, such as a neutron star or a stellar-mass black hole, spirals into a supermassive black hole found at the center of galaxies \cite{pan_formation_2021}. The detection of \acp{EMRI} holds significant scientific objectives \cite{amaro-seoane_laser_2017}. Firstly, they provide a unique probe into the spacetime geometry around supermassive black holes, allowing for precise tests of \ac{GR} in strong-field regimes \cite{babak_science_2017}. Secondly, \acp{EMRI} can offer insights into the evolution and demographics of compact objects in galactic centers, shedding light on the formation and growth of supermassive black holes over cosmic time \cite{babak_science_2017}.

From a technical perspective, detecting \acp{EMRI} poses substantial
challenges. The \ac{GW} signals produced by EMRIs are weak and buried in the
noise of space-based \ac{GW} detectors. Their waveform patterns are intricate due to the complex interplay of relativistic effects, making them difficult to model accurately \cite{gair_prospects_2017}. Furthermore, the long duration of
\ac{EMRI} signals demands efficient and robust data analysis techniques to comb
through vast amounts of data. Advanced computational methods, including machine
learning and deep neural networks, are being explored to enhance the detection
capabilities, such as
\citet{zhang_detecting_2022,zhao_dilated_2024,Yun_detecting_2023} detecting
EMRI signals using \acp{CNN}, achieving rapid detection of \acp{EMRI} with high
accuracy in time domain, frequency domain, and time-frequency domain. These
advancements are poised to play a pivotal role in the forthcoming era of
\ac{GW} astronomy, where space-based detectors like \ac{LISA}
\cite{amaro-seoane_laser_2017}, Taiji \cite{hu_taiji_2017,ren_taiji_2023}, and
TianQin \cite{luo_tianqin_2016} will offer unprecedented observational
capabilities for EMRI detection.

\subsection{GB}
\label{sec:detection_gb}

GW studies have revealed a range of astrophysical sources, each with distinct signal characteristics and associated detection challenges. Among these, continuous GWs emitted by \acp{GB} or isolated neutron stars—offer valuable insights into the dynamics of compact objects. In this section, we review the characteristics of continuous signals from galactic binaries, focusing on detection strategies and the innovative techniques employed to resolve these overlapping signals from noisy data.

Continuous \acp{GW}, unlike their transient counterparts, persist over extended
observation periods \cite{littenberg_global_2020}. These waves, often emanating from sources like rapidly rotating neutron stars, present a distinct challenge due to their weak amplitude \cite{littenberg_detection_2011}. Traditional data analysis methods often grapple with the intricacies of these continuous signals \cite{littenberg_global_2020}. However, the advent of \ac{DL} has ushered in a new era of possibilities. Neural networks, with their ability to learn
intricate patterns from vast amounts of data, have shown promise in detecting
and analyzing these weak, continuous \ac{GW} signals \cite{dreisigacker_searches_2020}. Several \ac{DL} models
have been proposed, aiming to enhance the detection capabilities by leveraging
the power of convolutional and recurrent architectures
\cite{dreissigacker_deep-learning_2019-1,dreissigacker_deep-learning_2020,dreisigacker_searches_2020,bayley_robust_2020,beheshtipour_deep_2021,yamamoto_use_2021}.
Although current pipelines for continuous signals rely predominantly on traditional signal processing methods, early studies suggest that deep learning could enhance the speed of these searches \cite{beheshtipour_deep_2020}. Ongoing research continues to assess the potential of these methods to complement traditional techniques in the pursuit of continuous GW detection \cite{joshi_large-kernel_2024,tenorio_one-stop_2024}.

\ac{LISA} is set to revolutionize our understanding of \acp{GW} from space.
However, one of the inherent challenges \ac{LISA} faces is the confusion noise.
This noise arises from the superposition of countless unresolved sources,
primarily binary systems, creating a cacophony that can mask potential signals.
Addressing this requires sophisticated data analysis strategies and a deep
understanding of the galactic \ac{GW} foreground.

\Ac{GP} have emerged as a powerful tool in the realm of \ac{GW} research, particularly for \ac{GB} GW separation \cite{falxa_adaptive_2023}. By leveraging the non-parametric nature of \acp{GP}, researchers can model the statistical properties of \acp{GB}, effectively separating it from potential signals \cite{strub_bayesian_2022}. This technique holds promise, especially in scenarios with overlapping signals or in the presence of non-Gaussian noise, offering a robust method to extract meaningful information from the data \cite{strub_global_2024}.

\section{Parameter Estimation}
\label{sec:pe}

\begin{figure*}
	\centering
	\includegraphics[width=0.9\textwidth]{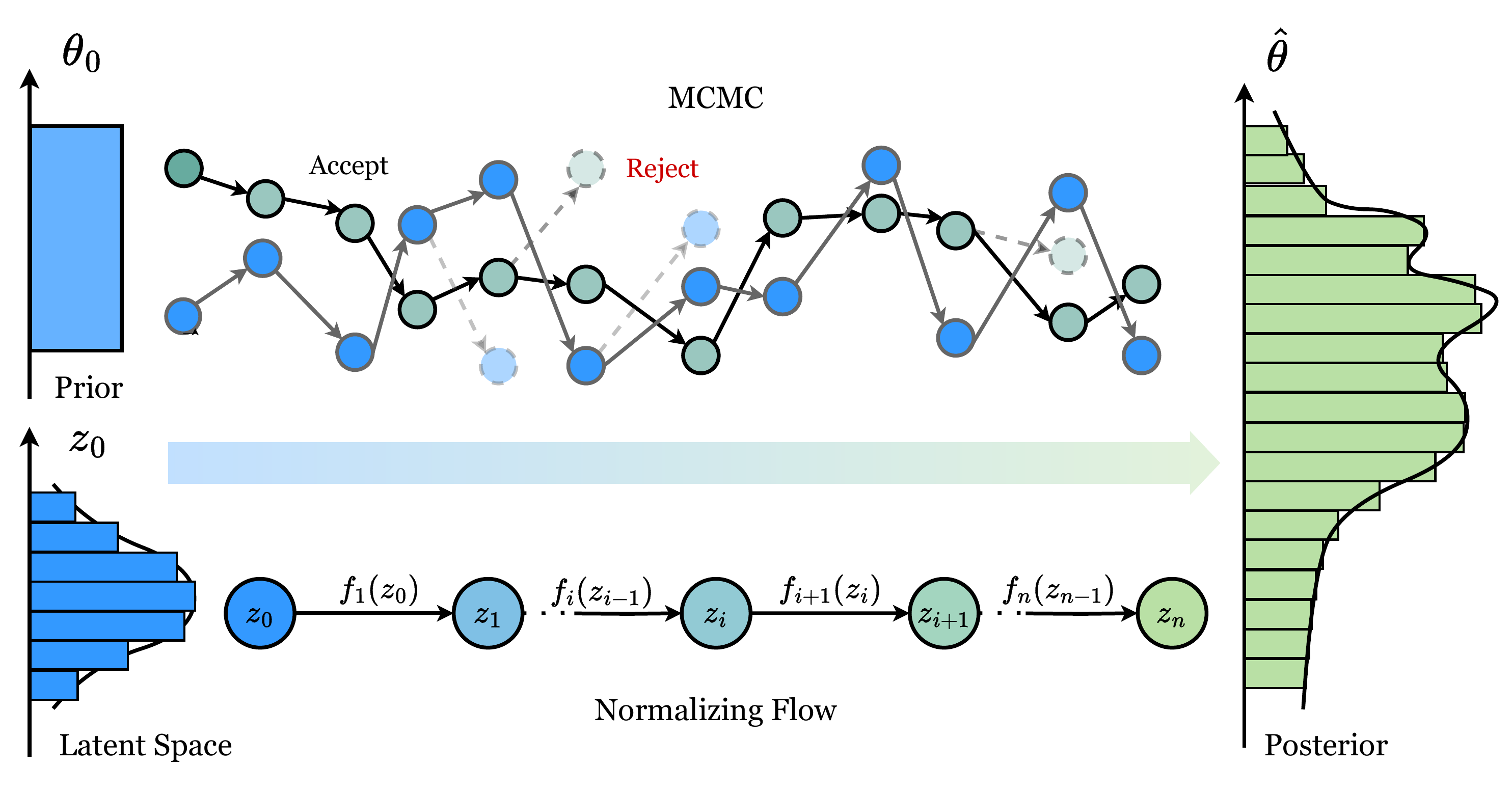}
	\caption{\textbf{Flowchart of the MCMC algorithm and normalizing flow.} The top portion of the diagram outlines the step-by-step process of the Metropolis-Hastings (MH) algorithm, which starts from an initial parameter value \(\theta_0\) (representing the prior) and iteratively updates proposals until converging to the final posterior distribution, \(\hat{\theta}\) (see, e.g., \cite{chib_understanding_1995}). The lower portion illustrates the normalizing flow approach, where a latent variable \(z_0\) is sampled from a standard normal distribution \(\mathcal{N}(0,1)\) and then transformed through a sequence of invertible mappings to yield the same posterior \(\hat{\theta}\) (see \cite{rezende_variational_2015}). This method significantly enhances computational efficiency compared to traditional MCMC techniques \cite{rezende_variational_2015}. Adapted from Ref. \cite{lee_metamodel_2015}.}
	\label{fig:mcmc}
\end{figure*}

The extraction of astrophysical source properties from observed \ac{GW}
signals, a process integral to \ac{GW} astronomy, is known as \ac{PE}. At the
heart of \ac{PE} lies the \ac{MCMC} algorithm, a statistical method that has
been instrumental in the field \cite{christensen_markov_1998}. \ac{MCMC}
operates on the principle of building Markov chains, which are sequences of
random samples that, over time, approximate the desired distribution of the
parameters being estimated (see \Cref{fig:mcmc} for detail). This method is
particularly effective in exploring high-dimensional parameter spaces, making
it widely used in \ac{GWDA} \cite{speagle_conceptual_2020}.

Despite the effectiveness of \ac{MCMC} in providing detailed insights into
complex data sets, it comes with its own set of challenges. The primary
concerns are its computational intensity and occasional struggles with
convergence, especially in scenarios involving intricate and voluminous data.
Moreover, global fitting—which aims to jointly estimate parameters for all overlapping signals in extremely high-dimensional spaces—is particularly challenging for space-based missions and third-generation ground-based detectors, significantly increasing the computational demands and speed requirements for parameter estimation \cite{littenberg_prototype_2023}.
These limitations become increasingly significant as the complexity of \ac{GW}
data escalates and the demand for rapid analysis grows
\cite{speagle_conceptual_2020}.

In response to these challenges, the field of gravitational-wave astronomy has begun to explore deep learning models as a complementary approach to traditional parameter estimation methods such as MCMC. Several studies have demonstrated that machine learning-based methods can offer potential improvements in processing large datasets and reducing computational time \cite{dax_real-time_2021,dax_flow_2023,dax2022group}. These methods are still under active development, and further research is needed to validate and refine them for operational use.

\subsection{Point Estimation}
\label{sec:pe_point}

\paragraph{Fast PE by MLP:}
One of the initial forays into \ac{DL} for \ac{PE} involved the use of
Multi-Layer Perceptrons (MLP). These architectures, designed to provide a
direct ``point estimate'' for each parameter, have the advantage of speed.
Capable of delivering estimates in near-real-time. Although MLPs are not currently a standard component of the production pipelines for multimessenger searches, early studies suggest that they could potentially be valuable in scenarios where rapid response is required \cite{krastev_detection_2021}.

\paragraph{Localization:}
Beyond the realm of intrinsic parameters, DL models have also been proposed to address extrinsic parameters. For instance, exploratory studies have suggested that DL-based approaches could be used to estimate both sky location and distance, potentially providing rapid source localization estimates for future multi-messenger observations \cite{sasaoka_localization_2022}.

\subsection{Posterior Distribution}
\label{sec:pe_dist}

While point estimates are invaluable for quick insights, a deeper understanding
of source properties necessitates the exploration of the full posterior
distribution. To this end, several \ac{DL} models have been proposed:

\paragraph{Flow-based Models:}
Flow-based architectures, which are known for their ability to model complex
distributions (\Cref{fig:mcmc}), have been a popular choice for density
estimation and posteriors. \citet{green_gravitational-wave_2020} and
\citet{shen_statistically-informed_2022-2} first demonstrated the efficacy of
these models in \ac{GW} \ac{PE} by proposing toy models for the low-dimensional
parameter sub-space of \ac{BBH}, then \citet{green_complete_2021} extended the
work to the full parameter space of \ac{BBH} and evaluated on GW150914.
Subsequently, \citet{wang_sampling_2022-1} accelerated the convergence speed by
intergrating domain knowledge into prior distribution. Recently,
\citet{dax_real-time_2021,dax2022group,dax_neural_2023,wildberger_adapting_2023-2,williams_importance_2023-1,williams_nested_2021-1,bhardwaj_peregrine_2023}
have further refined these models, showcasing their performance on a variety of
\ac{GW} events. Wong \textit{et al.}
\cite{wong_constraining_2021-2,wong_automated_2022,wong_fast_2023} developed a
\texttt{Jax}-based \cite{jax2018github} framework for rapid \ac{PE}. The ability of these models to
capture the nuances of the posterior distribution, especially in scenarios with
complex degeneracy, has been a key factor in their success. Furthermore, the
computational efficiency of these models, coupled with their ability to
generate samples, has potential in addressing the challenges of
traditional \ac{PE} methods.

\paragraph{Conditional Variational Autoencoders (CVAE):}
Another promising avenue is the use of CVAE. As demonstrated by
\citet{gabbard_bayesian_2022} and \citet{green_gravitational-wave_2020}, CVAEs,
by conditioning on observed data, can generate samples that align closely with
the true posterior distribution, offering a probabilistic perspective on source
parameters.

\paragraph{Gaussian Processes (GP):}
For specialized sources such as \acp{GB} and \acp{EMRI}, \ac{GP} have been
employed, as seen in \cite{falxa_adaptive_2023,liu_improving_2023}. Tailored to
the unique signatures of these sources, \acp{GP} provides refined posterior
estimates, enriching our comprehension of these intriguing astrophysical
systems.

\begin{table}[htbp]
	\begin{center}
		\caption{\textbf{Comparative analysis of AI techniques for GW parameter estimation.} This table offers a comparison of various AI methods in GW parameter estimation, highlighting not only their effectiveness as measured by the Jensen-Shannon divergence (JS div.) \cite{nielsen_jensenshannon_2019} against traditional MCMC results but also detailing the dimensionality of the parameter space each model is capable of sampling.}
		\label{tab:pe}%
		\begin{tabular}{@{}lccc@{}}
			\toprule
			\hline
			\textbf{Paper}                        & \textbf{Dim} & \textbf{Model} & \textbf{JS div.}   \\
			\hline\hline
			\citet{green_gravitational-wave_2020} & 5D           & CVAE           & --                 \\
			\citet{green_complete_2021};          & 15D          & nflow          & --                 \\
			\citet{dax_real-time_2021}            & 15D          & Dingo          & $2.2\times10^{-3}$ \\
			\citet{gabbard_bayesian_2022}         & 15D          & CVAE           & $\sim 0.1$         \\
			\citet{dax_neural_2023}               & 15D          & Dingo-IS       & $5\times10^{-4}$   \\
			\citet{langendorff_normalizing_2023}  & 30D          & nflow          & --                 \\
			\hline
			\bottomrule
		\end{tabular}
	\end{center}
\end{table}

\section{AI for GW Science}
\label{sec:science}

\Ac{GW} astronomy has opened a new window to the universe, allowing us to probe extreme astrophysical and cosmological phenomena. With the increasing complexity and volume of \ac{GW} data, \ac{AI} has emerged as a powerful tool to address various challenges and unlock new scientific potential in the field.

\subsection{Fundamental Physics}
\label{sec:science_phys}

\begin{figure}[t!]
	\centering
	\includegraphics[width=0.45\textwidth]{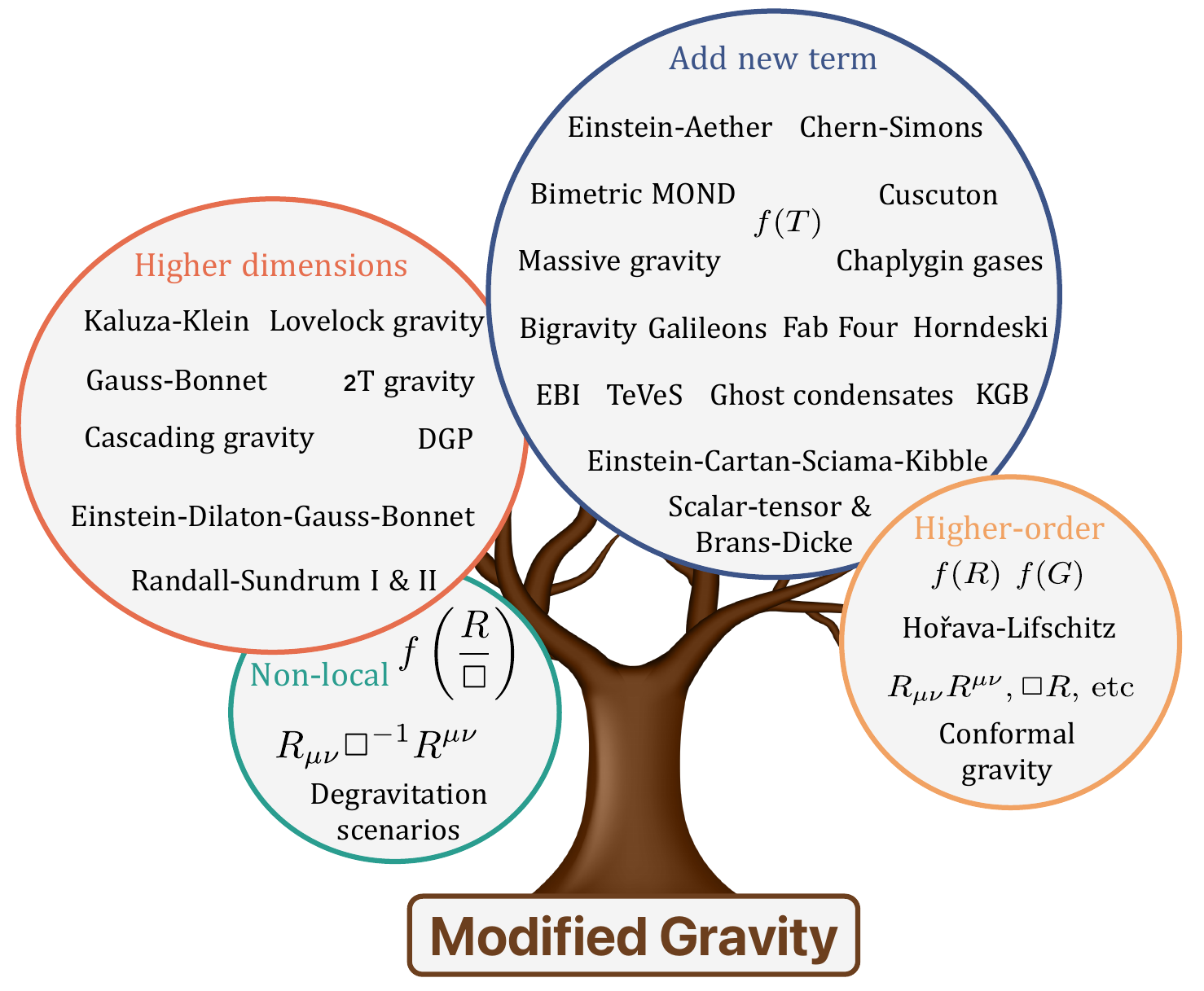}
	\caption{\textbf{Tree diagram of modified gravity theories.} This diagram illustrates the hierarchical structure and interrelationships among various modified gravity theories. Branching from the root concept, each pathway represents a distinct theoretical development, offering insights into alternative explanations of gravitational phenomena beyond general relativity. Adapted from Ref. \cite{bull_beyond_2016}.}
	\label{fig:modgrav}
\end{figure}

\Acp{GW}, ripples in the fabric of spacetime, have emerged as a revolutionary tool for astrophysical exploration. These waves, produced by cataclysmic events such as the merger of black holes or neutron stars, provide a pristine medium to probe the universe's most enigmatic phenomena \cite{bian_gravitational-wave_2021}. Unlike electromagnetic radiation, which can be obscured or altered by intervening matter, \acp{GW} travels undisturbed, offering a direct and unadulterated glimpse into their sources. The detection of \acp{GW} has opened a new avenue to explore the strong-field regime of \ac{GR} \cite{arun_new_2022}. Traditional electromagnetic observations, while invaluable, often fall short in this domain, especially when it comes to events like black hole mergers. \Ac{GW} detections, on the other hand, allow scientists to test Einstein's theory under extreme conditions, shedding light on the intricate dance of massive celestial bodies and the spacetime they warp \cite{arun_new_2022}.

The promise of space-based observatories, particularly \ac{LISA}, is immense.
Positioned far from Earth's noisy environment, \ac{LISA} aims to access the
low-frequency \ac{GW} spectrum \cite{bailes_gravitational-wave_2021}. This
capability is anticipated to unveil events like supermassive black hole
mergers, which have remained elusive to ground-based detectors, offering a new
observational window into the dynamics of galactic centers
\cite{barausse_massive_2020}. The \ac{GW} community eagerly anticipates the
advent of third-generation observatories, such as the \ac{ET}. These
state-of-the-art facilities promise enhanced sensitivity and a broader
frequency range \cite{maggiore_science_2020}. With these advancements,
scientists expect to detect a diverse array of sources, from the dramatic death
throes of massive stars in core-collapse supernovae to potential signals from
exotic compact objects. The enriched catalog of detections will undoubtedly
deepen our understanding of the universe's violent processes.

While \ac{GR} has withstood a century of scrutiny, \ac{GW} observations offer a
unique platform to test its predictions and probe potential deviations. The
precision of these observations might reveal subtle signatures hinting at
alternative theories of gravity, pushing the boundaries of our current
understanding and opening doors to new realms of physics (\Cref{fig:modgrav}).
The synergy of advanced detector networks has been instrumental in refining
parameter estimation techniques. As detectors' sensitivity improves, so does
the accuracy with which we can determine the properties of \ac{GW} sources.
This precision not only allows for a more detailed understanding of individual
events but might also offer subtle hints of new physics lurking in the shadows.
Among the most tantalizing prospects in \ac{GW} science is the detection of the
\ac{SGWB}. This omnipresent hum, a relic from the early universe, holds the
potential to offer insights into primordial processes and interactions,
painting a picture of the cosmos's infancy. Interdisciplinary efforts, merging
the expertise of astrophysicists, general relativists, and data scientists,
promise to usher in a new era of \ac{GW} science. As technology advances and
our observational capabilities expand, so too will our understanding of the
cosmos, revealing its mysteries one \ac{GW} at a time.

\subsection{Cosmology}
\label{sec:science_cosmos}
\Acp{GW} offer a unique lens to probe the cosmos, and their potential in cosmological studies is gradually being realized. One pressing issue in cosmology is the Hubble tension---the discrepancy between the Hubble constant values derived from cosmic microwave background radiation and those from local distance ladder measurements \cite{freedman_measurements_2021}. While traditional methods have presented conflicting results, \ac{AI}-driven analyses of \ac{GW} data could provide an independent and precise measurement of the Hubble constant \cite{the_ligo_scientific_collaboration_and_the_virgo_collaboration_gravitational-wave_2017}. By analyzing the \ac{GW} signals from binary mergers, \ac{AI} can help refine our understanding of the universe's expansion rate, offering a potential resolution to the Hubble tension. This synergy between \ac{GW} astronomy and \ac{AI} not only underscores the interdisciplinary nature of modern astrophysics but also promises to address some of the most perplexing challenges in the field \cite{stachurski_cosmological_2023}.

\subsection{Astrophysics}
\label{sec:science_astroph}

\Acp{GW} emanate from some of the most violent and energetic processes in the universe, making them invaluable tools for astrophysical studies \cite{abbott_gw170817_2017}. Neutron stars, the remnants of massive stellar explosions, are laboratories for extreme physics. Their \ac{EoS} remains one of the outstanding puzzles in astrophysics \cite{abbott_gw190425_2020}. \ac{AI}-driven methods can assist in classifying different \ac{EoS} models based on the \ac{GW} signals emitted during neutron star mergers \cite{fujimoto_mapping_2020,edwards_classifying_2021,morawski_neural_2020,soma_reconstructing_2023}. Furthermore, \ac{CCSN} are cataclysmic events marking the end of a massive star's life. The \ac{EoS} governing the processes inside these explosive events can be probed using \acp{GW} \cite{saiz-perez_classification_2022}. \ac{AI} can aid in deciphering the intricate signals from \ac{CCSN}, providing insights into the underlying physics \cite{portilla_deep_2021}. As the number of \ac{GW} detections grows, there's an increasing interest in understanding the population properties of the sources, be it \acp{BBH}, \acp{BNS}, or other exotic objects \cite{qiu_deep_2023}. \ac{AI} can assist in population synthesis studies, helping to unravel the formation and evolution histories of these compact objects \cite{wong_automated_2022}.

\section{Discussion}
\label{sec:discussion}

As the application of \ac{DL} techniques to \ac{GW} analysis matures, several
key themes and challenges emerge \cite{benedetto_ai_2023}. This section delves into a meta-analysis of the current landscape, discusses the prospects of waveform forecasting \cite{shi_compact_2024}, highlights the importance of gap imputation \cite{blelly_2021}, and underscores the significance of multi-modality \cite{cuoco_multimodal_2021} and interpretability \cite{khan_interpretable_2022-1} in \ac{DL} models.

\subsection{Meta-Analysis}
\label{sec:discussion_meta}

The intersection of \ac{GW} astronomy and \ac{DL} has generated a growing body of research. Our meta-analysis, presented in \Cref{fig:3} and \Cref{fig:4}, provides a quantitative overview of publication trends and the evolution of training dataset sizes within this field.
\Cref{fig:3} illustrates two key trends: the increase in average training dataset size over time and the rising number of publications—both peer-reviewed articles and preprints on platforms like arXiv. This analysis highlights how the availability of large datasets is associated with a corresponding increase in research output, which in turn supports the development of DL models for GW applications.
Each data point in these figures marks progress in the field, reflecting improvements in detector technology and data processing methods, as well as the expansion of the research community. In \Cref{fig:4}, the breakdown into subdomains—such as waveform modeling, signal detection, and parameter estimation—demonstrates the diversity of research efforts and their interconnections within the GW and AI.

While our meta-analysis offers a retrospective view of the past decade, the trends observed may also help inform future research directions and resource allocation. We acknowledge that these figures represent a limited perspective of a complex field; however, they serve as a useful overview of the evolving landscape at the intersection of GW astronomy and DL.

\subsection{Waveform Forecasting}
\label{sec:discussion_forecast}

\begin{figure}
	\centering
	\includegraphics[width=0.45\textwidth]{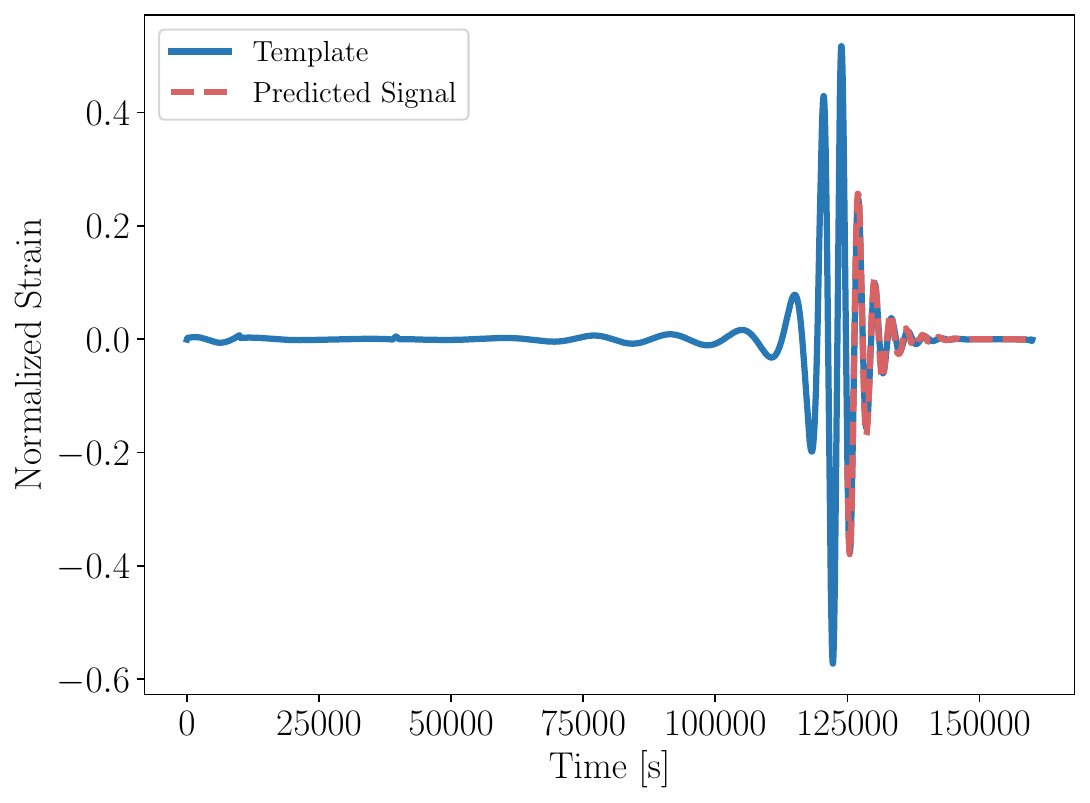}
	\caption{\textbf{GW signal forecasting.} This figure illustrates the forecasting of the GW waveform from MBHB. The blue line represents the template, and the red line represents the forecasted signal.}
	\label{fig:forecast}
\end{figure}

\ac{GW} astronomy has ushered in a new era of understanding the universe, with waveform modeling playing a pivotal role in deciphering the signals from astrophysical cataclysms. Waveform forecasting, a subset of this modeling, focuses on predicting the evolution of these waveforms (\Cref{fig:forecast}), especially in scenarios where only a fragment of the waveform is known or computationally feasible to generate.

The importance of efficient waveform modeling cannot be overstated. Traditional
waveform modeling, particularly for systems exhibiting higher modes or
precession, demands significant computational resources
\cite{mihaylov_pyseobnr_2023}. The ability to forecast a waveform, rather than
compute it in its entirety, offers a promising avenue to mitigate these
computational challenges \cite{nie_time_2022}. As the \ac{GW} community moves
towards real-time detections, the need for rapid template generation becomes
essential \cite{abbott_low-latency_2019}. Waveform forecasting can expedite
this process, ensuring that potential \ac{GW} events are not missed during live
observations. Some astrophysical systems present intricate waveforms that are
computationally intensive to model \cite{katz_fastemriwaveforms_2021}.
Forecasting provides a mechanism to capture the essence of these systems,
offering a balance between accuracy and computational feasibility.

While traditional methods have laid the groundwork for waveform forecasting,
they come with inherent limitations, especially when dealing with complex
astrophysical systems \cite{durbin_time_2012}. Enter \ac{DL}. Drawing
inspiration from its successes in other domains, particularly sequence
forecasting, \ac{DL} holds the promise of revolutionizing waveform forecasting
in \ac{GW} astronomy \cite{gu_efficiently_2022}. The potential benefits include
enhanced accuracy, reduced computational overhead, and the ability to handle a
broader range of astrophysical systems.

However, the journey of integrating \ac{DL} into waveform forecasting is not
without challenges. The complexity of \ac{GW} signals, the omnipresent detector
noise, and the need for vast training datasets to train robust models are but a
few of the hurdles to overcome \cite{shi_compact_2024}.

In conclusion, as we stand on the cusp of a new frontier in \ac{GW} astronomy,
the symbiosis between \ac{DL} and waveform forecasting is poised to play a
transformative role. Collaborative efforts between the \ac{AI} and \ac{GW}
communities will be instrumental in navigating this exciting journey ahead.

\subsection{Gap Imputation}
\label{sec:discussion_gap}

\begin{figure}
	\centering
	\includegraphics[width=0.45\textwidth]{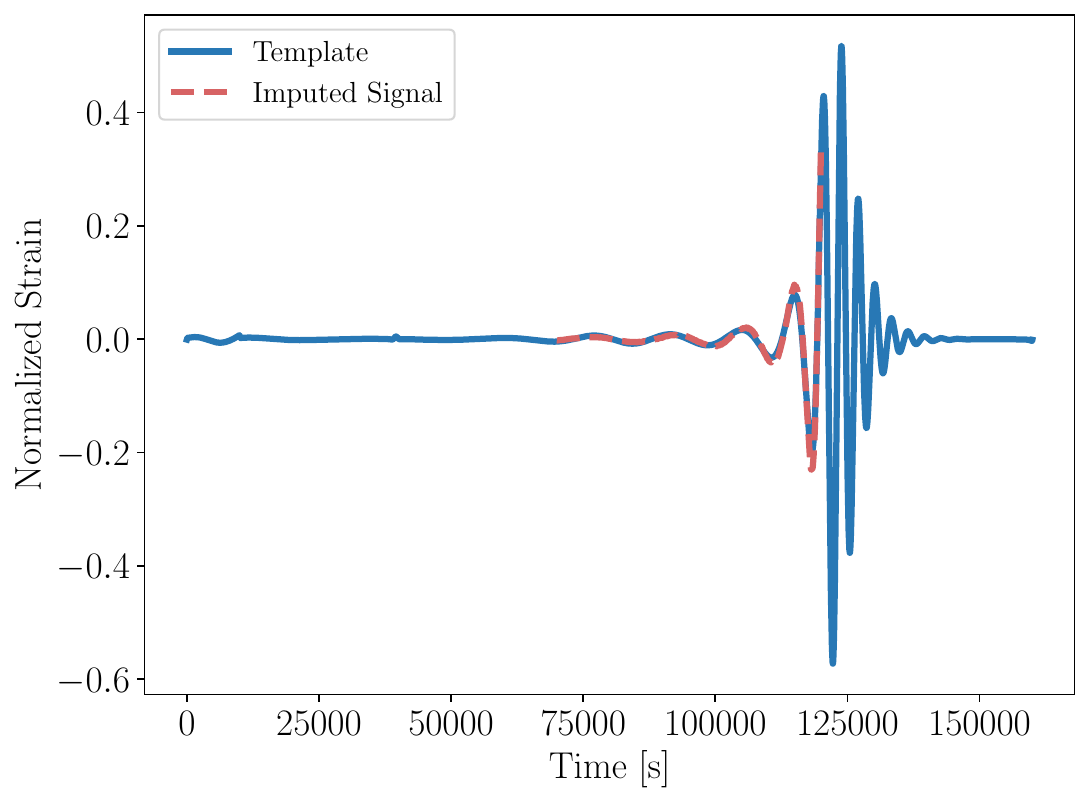}
	\caption{\textbf{GW signal imputation.} This figure illustrates the inpainting of the MBHB GW data gap. The blue line represents the template, and the red line represents the impainted signal.}
	\label{fig:gap}
\end{figure}

In the realm of time series data analysis, especially when dealing with time
series data, the presence of gaps or missing data points can pose significant
challenges \cite{lin_deep_2022}. These gaps can arise due to a myriad of
reasons, from instrumental downtime to environmental disturbances or even data
transmission issues \cite{diener_interspeech_2022}. \ac{GW} data, with its
intricate patterns and crucial reliance on continuity, is no exception to this
challenge \cite{ldc2b_baghi_2022}.

\ac{GW} observations, by their nature, require continuous and uninterrupted data streams for accurate data analysis \cite{baghi_gravitational-wave_2019}. Any gaps in this data can lead to biases in parameter estimation (\Cref{fig:gap}) \cite{blelly_2021}. Traditional methods employed in other fields to address missing data, such as linear interpolation or mean imputation, often fall short when applied to the complexities of \ac{GW} signals \cite{blackman_sparse_2014}. These methods, while simple, may not capture the intricate patterns and dependencies inherent in \ac{GW} data, leading to inaccuracies \cite{blelly_sparsity_2020}.

\ac{DL} is a paradigm that has shown remarkable success in various domains, including gap imputation in time series data \cite{emmanuel_survey_2021}.
\ac{DL} models, with their ability to learn and capture intricate patterns in
data, hold significant promise for addressing the gap imputation challenge in
\ac{GW} data \cite{li_end--end_2022}. The potential benefits are manifold, from
improved accuracy in imputation to the capability to handle large gaps that
traditional methods might struggle with. However, the application of \ac{DL} to
\ac{GW} data for gap imputation is not without its challenges. The
non-stationary nature of \ac{GW} data, combined with the noise,
necessitates careful consideration and design of \ac{DL} models \cite{le_missing_2024}. Additionally,
it is essential to ensure that the integration of imputed data does not lead to
biases or inaccuracies in further analytical processes \cite{castelli_extraction_2024}.

In conclusion, while \ac{DL} offers a promising avenue for gap imputation in
\ac{GW} data, the field is ripe for research and exploration. Collaborative
efforts between the \ac{AI} and \ac{GW} communities could pave the way for
innovative solutions, ensuring that our observations of the universe remain as
accurate and uninterrupted as possible.

\subsection{Multi-Modality}
\label{sec:discussion_multimodal}

\begin{figure}[h!]
	\centering
	\includegraphics[width=0.45\textwidth]{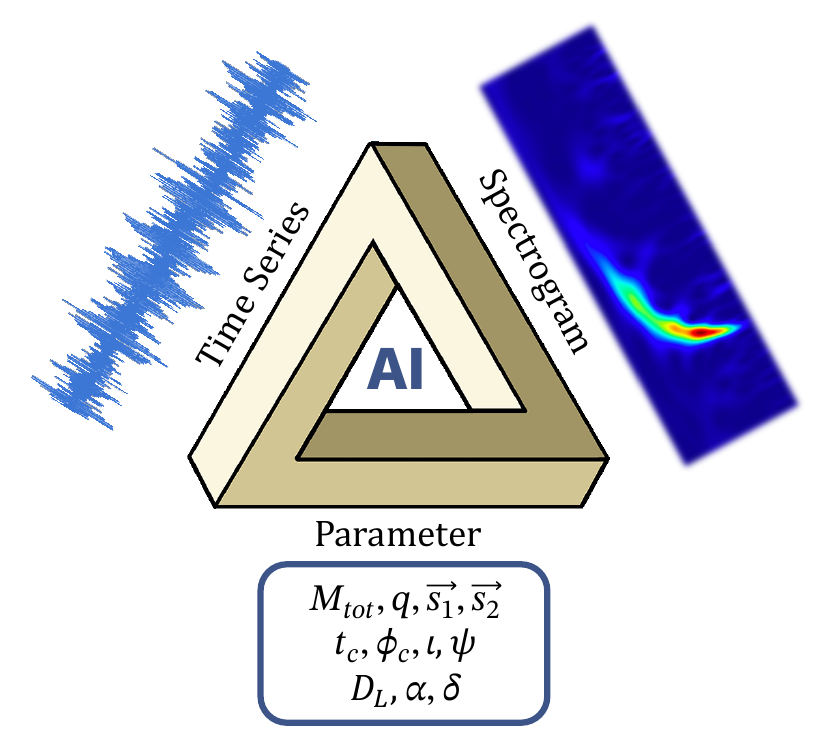}
	\caption{\textbf{Multi-modal representation of GW data for large AI model training.} This figure showcases a comprehensive analysis framework combining time series data, spectrogram visualizations, and physical parameters. The time series plot captures the temporal dynamics of the GW signals, while the spectrogram provides a frequency-based perspective. The inclusion of physical parameters, such as mass and spin, offers a deeper understanding of the intrinsic physical properties.}
	\label{fig:modality}
\end{figure}

The realm of \ac{AI}, particularly in the domain of \ac{DL}, has witnessed a
surge in the exploration and application of multi-modal techniques
\cite{rani_applying_2022,pei_review_2023}. These techniques harness information
from multiple data sources or modalities, aiming to provide a more
comprehensive understanding of the underlying phenomena. In many \ac{AI}
applications, multi-modality has proven to be transformative. For instance, in
medical imaging, combining visual data from MRI scans with textual patient
records has enhanced diagnostic accuracy and predictive modeling
\cite{korot_code-free_2021}. Similarly, in natural language processing, the
fusion of textual, auditory, and visual cues in multi-modal sentiment analysis
models has led to more nuanced and context-aware interpretations
\cite{zhao_adaptive_2021}.

\
\ac{GW} astronomy, while primarily reliant on signal data, has the potential to benefit from a multi-modal approach. As a starting point, one can combine raw time series data with their time-frequency representations—such as spectrograms—to capture complementary features. This initial idea, though relatively simple, forms the basis for more sophisticated multi-modal analyses (\Cref{fig:modality}). Consider the richness of information available: beyond the \ac{GW} signals, there are electromagnetic signals, neutrino observations, and more \cite{meszaros_multi-messenger_2019}. Each modality offers a unique perspective on astrophysical events. While current \ac{GW} analyses have not yet integrated such diverse data streams, the success of multi-modal techniques in other \ac{AI} domains suggests potential avenues for exploration. One can envision a future where \ac{GW} detections are enhanced by concurrent observations from electromagnetic or neutrino observatories \cite{abbott_multi-messenger_2017}. \ac{DL} models, adept at handling multi-modal data, could be trained to extract features from each modality and then fuse these features to improve event characterization, source localization, and parameter estimation \cite{hoy_rapid_2024}.

However, the integration of multi-modal data in \ac{GW} analysis is not without
challenges. The synchronization of data streams, the handling of disparate data
resolutions and formats, and the development of fusion techniques tailored to
the specificities of \ac{GW} events are all areas that would require meticulous
research and innovation \cite{cuoco_multimodal_2021}. Yet, the potential rewards are significant. A successful multi-modal approach could provide a more holistic view of
astrophysical events, bridging gaps in our understanding and opening new
frontiers in \ac{GW} astronomy \cite{hoy_rapid_2024}.

\subsection{Interpretability}
\label{sec:discussion_interpret}

\begin{figure}[ht!]
	\centering
	\includegraphics[width=0.45\textwidth]{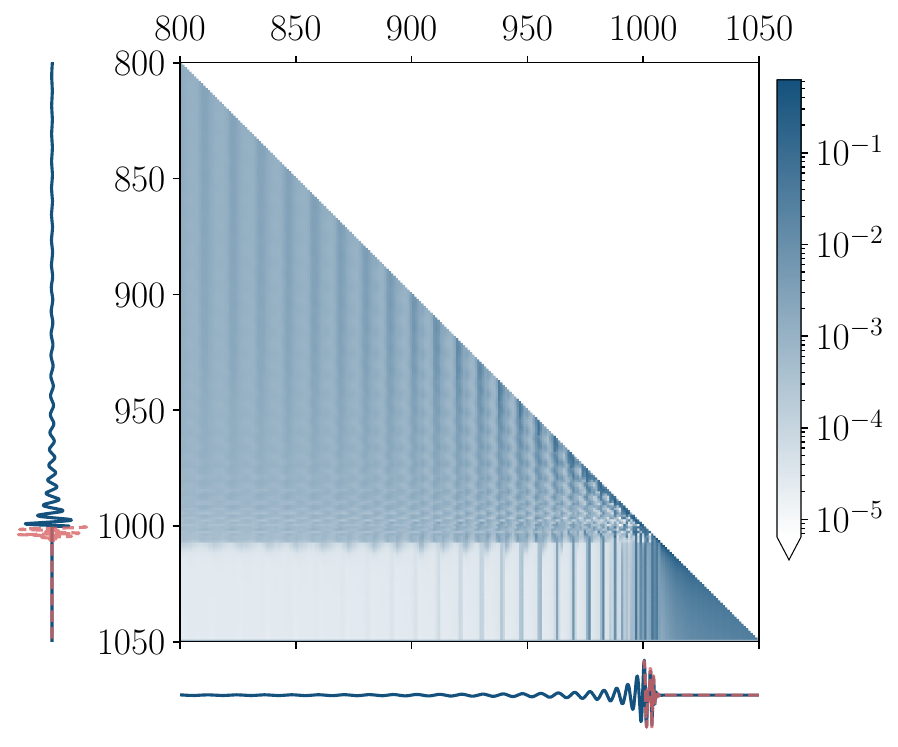}
	\caption{\textbf{Interpretability showcase of the pretrained large AI model in GW waveform prediction.} This figure features a colored mesh representing the attention map of the transformer model, which highlights the weights the model focuses on during analysis. The blue lines in the left and bottom panels show the input GW waveforms. The red line illustrates the waveform predicted by the model. The attention map demonstrates the model's capability for accurately modeling and forecasting gravitational wave signals. Adapted from Ref. \cite{shi_compact_2023}.}
	\label{fig:att_map}
\end{figure}

Interpretability in machine learning has garnered significant attention
\cite{lipton_mythos_2018}, especially when models are applied to complex
scientific domains \cite{doshi-velez_towards_2017}. In the realm of \ac{GW}
analysis, understanding the decision-making process of models is not just a
luxury but a necessity. This ensures that the predictions and insights derived
are both reliable and scientifically meaningful. \ac{GW} signals, with their
intricate patterns and the profound astrophysical phenomena they represent,
pose a unique challenge for \ac{DL} models \cite{khan_physics-inspired_2020}. While these models can achieve impressive performance metrics, deciphering their decision-making process in the context of such complex signals remains a formidable task \cite{khan_interpretable_2022-1}.

Recent works have made strides in bridging this interpretability gap. For
instance, \citet{khan_interpretable_2022-1} delved into creating models that
not only detect but also elucidate the characteristics of \ac{GW} signals.
Similarly, the studies by \citet{zhao_space-based_2023,shi_compact_2023} have
contributed to enhancing the transparency of models, especially Transformers,
ensuring that their predictions can be traced back to understandable reasoning.
An interpretable model in \ac{GW} analysis offers more than just reliable
predictions (see \Cref{fig:att_map}). It provides a window into the
astrophysical processes that generate these ripples in spacetime. By
understanding how a model discerns between different types of signals,
researchers can gain deeper insights into the astrophysical events behind these
waves, potentially unlocking new facets of our universe.

As the field progresses, there's a palpable need for more research dedicated to
enhancing interpretability. The fusion of traditional astrophysical knowledge
with transparent machine learning models holds the promise of richer, more
profound insights into the cosmos.

\begin{acknowledgments}
	The research was supported by the Peng Cheng Laboratory and by Peng Cheng Laboratory Cloud-Brain.
	This work was also supported in part by the National Key Research and Development Program of China Grant No.~2021YFC2203001 and in part by the NSFC (No.~11920101003 and No.~12021003). Z.C was supported by the ``Interdisciplinary Research Funds of Beijing Normal University" and CAS Project for Young Scientists in Basic Research YSBR-006.
\end{acknowledgments}

\bibliographystyle{myapsrev4-2_title}
\bibliography{references}

\end{document}